\documentclass[12pt,letterpaper]{JHEP3}

\usepackage{slashed}
\usepackage{amsmath}
\usepackage{amssymb}
\usepackage{axodraw}
\usepackage{bm}

\newcommand{\ba}{\begin{eqnarray}}
\newcommand{\ea}{\end{eqnarray}}
\newcommand{\rmi}[1]{{\mbox{\scriptsize #1}}}

\newcommand{\tr}{{\rm tr\,}}
\newcommand{\nn}{\nonumber \\}
\newcommand{\fr}[2]{{\frac{#1}{#2}\,}}

\renewcommand{\(}{\left(}
\renewcommand{\)}{\right)}

\newcommand{\e}{\epsilon}

\newcommand{\tfr}[2]{{\textstyle \frac{#1}{#2}\,}}

\def\sumint{\hbox{$\sum$}\!\!\!\!\!\!\!\int}

\renewcommand{\ln}{{\rm ln}}

\def\openone{\rlap 1\kern 0.22ex 1}

\def\sintpm{\hbox{$\sum$}\!\!\!\!\!\!\!\int_{\pm} \frac{d^4 k}{(2 \pi)^4}}
\def\sintp{\hbox{$\sum$}\!\!\!\!\!\!\!\int_{+}  \frac{d^4 k}{(2 \pi)^4}}
\def\sintm{\hbox{$\sum$}\!\!\!\!\!\!\!\int_{-}  \frac{d^4 k}{(2 \pi)^4}}

\newcommand{\real}{{\rm Re}}
\newcommand{\im}{{\rm Im}}
\newcommand{\Erg}{\mathcal E}

\def\sumint{\hbox{$\sum$}\!\!\!\!\!\!\!\int}

\preprint{CERN-PH-TH/2009-076 \\
TUW-09-06}
\title
    {
    \boldmath
    On the dispersion of fundamental particles in QCD and ${\mathcal N} =4$ Super Yang-Mills theory
    }
\author
    {%
    P.~M.~Chesler$^1$,
    A.~Gynther$^2$, and A.~Vuorinen$^{23}$
    \\
$^1$Department of Physics, University of Washington, Seattle, WA 98195--1560
$^2$Institut f\"ur Theoretische Physik, Technische Universit\"at Wien, Wiedner Hauptstr.\\
\hskip 0.05in 8-10, A-1040 Vienna, Austria \\
$^3$CERN, Physics Department, TH Unit, CH-1211 Geneva 23, Switzerland\\
\email{pchesler@u.washington.edu, gynthera@hep.itp.tuwien.ac.at,
aleksi.vuorinen@cern.ch}
}

\abstract {We study thermal corrections to the dispersion relations of massive fundamental particles immersed in weakly coupled non-Abelian plasmas. The cases covered include quarks in the QCD (quark-gluon) plasma, as well as ${\mathcal N}=2$ quarks and scalars in an ${\mathcal N}=4$ Super Yang-Mills plasma. We perform the calculations to leading order in a weak coupling expansion, and consider all mass scales of the fundamental fields, ranging from massless particles all the way to bare masses parametrically larger than the temperature.}


\begin{document}


\section{Introduction}

The goal of the paper at hand is to study weakly coupled deconfined gauge theories at finite temperature, and in particular to determine the leading order thermal corrections to the dispersion relations of charged fundamental representation particles. The dispersion relation provides the on-shell relationship between an excitation's energy and momentum, as well as its lifetime, and hence encodes valuable information about the nature of the quanta making up the theory. In this paper, we consider the weakly coupled limits of both QCD (with $N_l$ flavors or massless quarks) as well as ${\mathcal N} =4$ Super Yang-Mills theory (SYM). To these theories, we add one fundamental representation test particle with an arbitrary mass, which for QCD is a quark and for SYM an ${\mathcal N} =2$ hypermultiplet, containing one Dirac fermion and two complex scalars of the same mass. Our presentation will cover all mass scales of the fundamental particle, from the massless limit to fields with masses parametrically larger than the temperature.

The original motivation for our work arose from various studies of strongly coupled ${\mathcal N} =4$ SYM, where the dispersion relations of fundamental particles had been determined using the AdS/CFT correspondence (see \textit{e.g.~}Ref.~\cite{hky}). In the course of our project, we however noted the surprising absence of some of the corresponding results in the QCD literature --- in particular in the limit of heavy quarks, with $M\gtrsim T$ --- and therefore decided to include this theory in our presentation as well. Our paper thus contains a self-consistent study of the leading order (LO) thermal corrections to the dispersion relations of fundamental particles in both theories and for all mass scales. We note, however, that a subset of the results presented in Section 5 on QCD can be found from the literature, in particular in Refs.~\cite{weldon1,petit,pisa}. In addition, in the massless limit of QCD, even the NLO thermal corrections to the dispersion relation of a static quark have been recently determined \cite{Carrington:2008dw}.

For QCD with massless and light quarks at soft (${\mathcal O}(gT)$) momenta, our findings confirm the earlier results of Refs.~\cite{weldon1,petit,pisa}, which we furthermore observe to be closely analogous to the ones we obtain for fundamental quarks in ${\mathcal N}=4$ SYM. For the latter, one merely needs to change the value of the soft mass parameter appearing in the QCD result, but its form stays otherwise intact. Moving then on to SYM with either bare masses or three-momenta of order $T$, we find a simple dispersion relation $E=\sqrt{p^2+M^2+\delta M^2}$ for both quarks and scalars, of which the latter in fact obey this relation even in the soft regime. For SYM, the thermal mass shift $\delta M^2$ is independent of the momentum $p$, whereas in QCD, the dispersion relation is seen to have a non-trivial form for practically all bare masses and momenta.

In the latter part of our paper, we specialize to the case where the bare mass of the fundamental particle is parametrically larger than the temperature, $M \gtrsim T/g$. This is a particularly interesting regime to study, as it allows us to compare the weak coupling results obtained in SYM to the corresponding strong coupling calculations of Ref.~\cite{hky}. Regardless of the strength of the coupling, we expect to find finite corrections to the dispersion relation due to infrared (IR) physics, as thermal screening and damping effects of long wavelength degrees of freedom must change the energy stored in the fields sourced by the heavy fundamental particle. At weak coupling, the IR sensitivity of the dynamics requires us to use resummed perturbation theory in determining the dispersion relation, but we argue that the corresponding corrections can just as well be computed using macroscopic classical field theory, where thermal effects are taken into account through temperature dependent screening masses as well as permittivity and permeability tensors. To demonstrate this, we compute the leading order non-relativistic dispersion relation both using classical field theory and resummed perturbation theory. The two approaches are seen to lead to the same outcome, which is later compared to the strong coupling result.

The paper is organized as follows. In Section 2, we introduce our notation and conventions, write down the Lagrangians of the two theories, and explain in detail, how we intend to approach the task at hand. Sections 3, 4 and 5 on the other hand contain the bulk of our calculations, the two former dealing with ${\mathcal N} =4$ SYM theory and the last one with QCD. In these Sections, we assume the mass of the fundamental particle to be at most of the order of the temperature, while the non-relativistic limit of $M\gg T$ is left to Sec.~6. In Section 7, we then discuss the main implications of our results, as well as briefly draw conclusions. While most of the details of our calculations are presented in the main body of the text, we have decided to separate the lengthy computation of one specific integral, contributing to the self energy of an infinitely massive quark, into Appendix A. The treatment in the main text is however entirely self-consistent, as we present the entire classical calculation in Section 6. In addition, we list the results for some straightforwardly computable sum-integrals in Appendix B, but omit the derivations that can be found from several textbooks (see \textit{e.g.~}Ref.~\cite{kap}).

\section{Setup}

In field theory, the dispersion relation of an excitation is obtained by studying the location of the pole(s) of the corresponding Minkowski space two-point function. The location of the pole encodes information about both the kinematics and lifetime of the excitation, thus providing insight into its real time properties. It is, however, both possible and convenient to obtain the dispersion relation from time-ordered Euclidean correlation functions. As can be easily demonstrated by decomposing them in terms of spectral densities, time-ordered Euclidean two-point functions contain the same information as their real time counterparts, and in particular, upon analytic continuation to Minkowski space, have poles in the same locations \cite{kap,fw}.

The time-ordered Euclidean correlation functions, as well as their poles, in general have both real and imaginary parts, but one often finds that the imaginary parts are suppressed by higher powers of the coupling constant than the real ones. As the real part of the dispersion relation gives the on-shell relationship between the corresponding excitation's energy and momentum and the imaginary part its decay width, the fact that we only work to the leading perturbative order in the present paper implies that we will not be able to extract information on the widths.

Being solely interested in the finite temperature corrections to dispersion relations in this work, we always regularize our sum-integrals by subtracting away their (often divergent) zero temperature pieces. Due to only working to leading order in a weak coupling expansion, this simply amounts to defining the tree level mass we use in our equations as one already containing all zero-temperature quantum corrections, \textit{i.e.~}being the renormalized, physical mass parameter. Beyond this, we never need to worry about the details of renormalization.

\subsection{Notation and conventions}

Our notation is as follows.  We use capital letters $P$ to denote four-vectors, lower case bold ones $\bm p$ to denote three-vectors, and the usual lower case letters $p$ to denote the magnitude of the latter. Sum-integrals will, as usual, be written as ($d=4-2\epsilon$)
\ba
\sumint_{P/\{P\}} &\equiv& \Lambda^{2\epsilon}T\,\sum_n \int\fr{{\rm d}^{d-1}p}{(2\pi)^{d-1}},
\ea
and their finite parts, where the $T=0$ contribution has been subtracted off and $d$ subsequently set to $4$, as
\ba
\sumint_{\pm}\fr{{\rm d}^4p}{(2\pi)^4} &\equiv& \sumint_{P/\{P\}}-\Lambda^{2\epsilon}\int\fr{{\rm d}^{d}p}{(2\pi)^{d}}\mbox{\Huge{$\mid$}}_{\e\rightarrow 0}.
\ea
Here, the subscript $P$ and the sign $+$ refer to bosonic and $\{P\}$ as well as $-$ to fermionic Matsubara sums, with $p_0=2n\pi T$ and $p_0=(2n+1)\pi T$, $n\in \mathbb{Z}$, respectively. Unless otherwise stated, also the zero components of the external momenta in the graphs we consider will be assumed to take these values. The parameter $\Lambda$ denotes here the MS renormalization scheme scale parameter, but will never show up in the final results.

The analytic continuation to the Minkowski space is defined by $p_0 \rightarrow i \Erg + \delta$, where $\delta$ is an infinitesimal positive real number and $\Erg$ the Minkowski space energy. We denote the unperturbed dispersion relation by $E(p) = \sqrt{p^2 + M^2}$, and from Section 3 onwards will work exclusively with the Euclidean metric, $g_{\mu\nu}=\delta_{\mu\nu}$. Finally, a group theory constant that will be in frequent use in the following is the Casimir of the fundamental representation of SU$(N_c)$,
\ba
C_F\,\delta_{ij}&=&\(T^a T^a\)_{ij}\,=\,\fr{N_c^2-1}{2N_c}\delta_{ij}.
\ea

\subsection{The Lagrangians}

The two theories we study in this paper are ${\mathcal N} =4$ SYM theory coupled to one fundamental ${\mathcal N} =2$ hypermultiplet with mass $M$, as well as QCD with $N_l$ massless quarks and one test quark with mass $M$.\footnote{Due to the low perturbative order we are working in, the parameter $N_l$ will only be visible through its effects on the Debye mass $m_D$ of QCD.} They are defined through the following four-dimensional Lagrangians, which we give in Minkowski space with $-+++$ signature.

The field content of ${\mathcal N} =4$ SYM consists of one gauge field $A_{\mu}$, four Majorana fermions $\psi_i$ and three complex scalars $\phi_p$, while the additional $\mathcal N=2$ multiplet is composed of two complex scalars $\Phi_n$ and one Dirac fermion $\omega$. All $\mathcal N=4$ fields transform under the adjoint representation of the gauge group SU($N_c$), while the $\mathcal N=2$ sector transforms under the fundamental representation. The ${\mathcal N} =4$ fields are massless, and the ${\mathcal N} =4$ theory conformal, while the ${\mathcal N} =2$ fields each have mass $M$.

Following Ref.~\cite{yy}, we define $\phi_p = 1/\sqrt{2} \(X_p + i Y_p \)$, with $X_p$ and $Y_p$ hermitian, which allows us to write our Lagrangian in the form \cite{cv}
\ba
\label{lagrangian}
\mathcal L = \mathcal L_0 + \mathcal L_1 + \mathcal L_2,
\ea
with
{\setlength{\baselineskip}{1.4\baselineskip}
\ba
\label{l0}
\mathcal L_0&=&  -\tr \Big \{\frac{1}{2} F_{\mu \nu} F^{\mu \nu}+ \bar \psi_i \slashed{D} \psi_i +\(D X_{p}\)^{2}+\(D Y_{p}\)^{2} \Big \}  \nn
&-&(\Phi)_n^{\dagger} (-D^{2}+M^2) \Phi_n -\bar \omega (\slashed{D} +M)\omega,\\
\label{l3}
\mathcal L_1/g &=& \tr \Big \{-i \bar \psi_i \alpha_{ij}^{p} [X_p,\psi_j] + \bar \psi_i \gamma_5 \beta^{p} _{ij}[Y_p, \psi_j] \Big \} -\bar \omega
\left ( Y_1 - i \gamma_5 X_1 \right ) \omega  \nn
&+&2 \sqrt{2} \,\text{Im} \Big( - \bar \omega P_{+} \psi_1 \Phi_1 - (\Phi_2)^{\dagger} \bar \psi_1 P_{+} \omega +  (\Phi_1)^{\dagger} \bar \psi_2 P_{+} \omega
- \bar \omega P_{+} \psi_2 \Phi_2 \Big) \nn
&-&2 M (\Phi)^{\dagger}_{n} Y_{1} \Phi_{n},\\
\label{l4}
\mathcal L_2/g^2 &=&-\frac{1}{2} \tr \( i [\chi_A, \chi_B]\)^2+(-1)^{n}  (\Phi_n)^{\dagger} \( [\phi_2,\phi_2^{\dagger}] +
[\phi_3,\phi_3^{\dagger}] \) \Phi_n \nn
&-& 4 \text{Re} \( (\Phi_1)^{\dagger} [\phi_2,\phi_3] \Phi_2\)-\tfr{1}{2} \big | (-1)^{n} (\Phi_{n})^{\dagger} t_a  \Phi_{n} \big |^{2}
-2  \big | (\Phi_2)^{\dagger} t_a \Phi_1 \big |^{2} \nn
&-& (\Phi_n)^{\dagger} \{\phi_1,\phi_1^{\dagger} \} \Phi_{n}.
\ea
\par}
\noindent Here, $D$ denotes covariant derivatives in the appropriate representations of SU($N_c$), $\chi \equiv (X_1,Y_1,X_2,Y_2,X_3,Y_3)$, and a sum over repeated indices is implied. The matrices $\alpha^p$ and $\beta^p$ are given by
\begin{subequations}
\label{coefficients}
\begin{align}
    \alpha^1&=
    \begin{pmatrix}
        i\sigma_2 & 0    \\
        0 & i\sigma_2
    \end{pmatrix}   ,\;\;\;\;
    \alpha^2=
    \begin{pmatrix}
        0 & -\sigma_1   \\
        \sigma_1 & 0
    \end{pmatrix}   ,\;\;\;\;\;\;
    \alpha^3=
    \begin{pmatrix}
        0 & \sigma_3   \\
         -\sigma_3 & 0
    \end{pmatrix}   ,\;
                \\
    \beta^1&=
    \begin{pmatrix}
        -i\sigma_2 & 0   \\
        0 & i\sigma_2
    \end{pmatrix}   ,\;
    \beta^2=
    \begin{pmatrix}
        0 & -i\sigma_2    \\
        -i\sigma_2 & 0
    \end{pmatrix}   ,\;
    \beta^3=
    \begin{pmatrix}
        0 & \sigma_0  \\
         -\sigma_0 & 0
    \end{pmatrix}   \, ,
\end{align}
\end{subequations}
and they satisfy the algebra
\ba
\{\alpha^p,\alpha^q\} &=& \{\beta^p,\beta^q\} = -2 \delta^{pq} ,\nn
\big[\alpha^p,\beta^q\big] &=& 0.
\ea

Finally, for QCD the bare Lagrange density reads
\ba
{\cal L}_\rmi{QCD} & = &  -\fr12 \tr F_{\mu\nu} F^{\mu\nu} - \bar\psi_f\(\slashed{D}+M_f\)\psi_f,
\ea
where $F^{\mu \nu}$ is the field strength, $\psi_f$ are fundamental Dirac fermions and
$D$ is the gauge-covariant derivative. The flavor sum over $f$ includes $N_l$ massless flavors with $M_f=0$ and one test quark with $M_f=M$.

\section{Fundamental scalars in ${\mathcal N} =4$ SYM \label{scalarsec}}

\begin{FIGURE}[t]
{\centerline{\epsfxsize=15.0cm\epsfysize=3.5cm\epsfbox{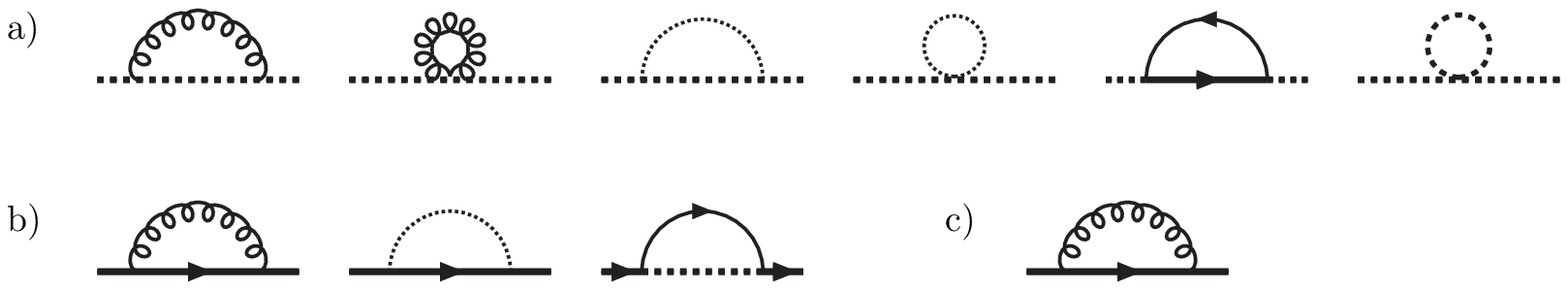}}

\caption[a]{The one-loop diagrams contributing to the self energy of a fundamental a) scalar and b) fermion in the ${\mathcal N}=4$ SYM theory, and c) a quark in QCD. The thick and thin dotted lines stand for the ${\mathcal N}=2$ and ${\mathcal N}=4$ scalars, respectively, and the thick and thin solid lines for the fundamental and ${\mathcal N}=4$ fermions. The wiggly line on the other hand denotes a gluon. \label{graphs1}}}
\end{FIGURE}

Let us begin by considering the dispersion relation of a fundamental ${\mathcal N} =2$ scalar in the ${\mathcal N} =4$ SYM theory. To this end, we note that the Euclidean scalar self energy is given by the sum of the graphs of Fig.~\ref{graphs1}.a and reads in the Feynman gauge
\ba
\label{scalar1}
\Pi(P) &=& 4 g^2 C_F \sintp \left ( \frac{1}{K^2} + \frac{1}{K^2 + M^2} - \frac{1}{2}\frac{P^2 + M^2}{(P+K)^2 + M^2} \frac{1}{K^2} \right )
\\ \nonumber \\ \nonumber
&-& 4 g^2 C_F \sintm \left ( \frac{1}{K^2} + \frac{1}{K^2 + M^2} - \frac{P^2 + M^2}{(P+K)^2 + M^2} \frac{1}{K^2} \right ).
\ea
The equation for the pole in the scalar propagator is then simply given by
\ba
\label{polex}
P^2 + M^2 + \Pi(P) &=& 0,
\ea
from which we wish to solve $p_0$. If $\Pi(P)$ is regular in the unperturbed on shell limit $P^2 = -M^2$, \textit{i.e.~}does not diverge there, then Eq.~(\ref{polex}) may be solved to leading order using this relation in the evaluation of $\Pi(P)$. With this in mind, we now consider the terms proportional to $P^2 + M^2$ in Eq.~(\ref{scalar1}), which contain the sum-integrals of Eq.~(\ref{B3}) of Appendix B.

To see how these functions behave in the unperturbed on shell limit, we analytically continue $p_0$ to real energies $E(p) = \sqrt{p^2 +M^2}$
by setting $p_0 = i E(p) +\delta$, with $\delta > 0$ . The logarithm appearing in the first term of Eq.~(\ref{B3}) now becomes
\ba
\log \left ( \frac{  (E( p)+p  )k + i (E( p) -k) \delta}{(E( p)+p  )k - i (E( p) +k) \delta} \
\frac{( E( p)-p )k - i (E(p) +k) \delta}{ (E( p)-p  )k + i (E( p)-k) \delta}
\right ),
\ea
which in fact vanishes in the $\delta \rightarrow 0$ limit for all non-zero $M$, and thus also in the $M\rightarrow 0$ limit. At the same time, the log in the second term of Eq.~(\ref{B3}) obtains the form
\ba
\log \left ( \frac{M^2 - E(k) E(p)+ p k  }{ M^2 - E( k) E(p)-p k  } \ \frac{M^2 + E( k) E( p)+ p k  }{M^2 + E( k) E( p)- p k } \right ) &=&
\log \(\frac{(k-p)^2}{(k+p)^2}\),\label{logid}
\ea
which obviously contains an integrable logarithmic singularity at $k=p$ but is otherwise regular. It follows that the integrals in Eq.~(\ref{scalar1}) are regular in the unperturbed on shell limit, and that the terms in $\Pi(P)$ proportional to $P^2 + M^2$ may be thus neglected when evaluating the function.

Throwing now away the terms discussed above and using Eq.~(\ref{B2}) from Appendix B to evaluate the remaining sum-integrals, we arrive at the $p$ independent result for the self energy,
\ba
\label{scalar}
\Pi(P^2=-M^2) &=&  \frac{4g^2 C_F}{\pi^2} \left \{ \int_{0}^{\infty} dk k \frac{e^{\beta k}}{e^{2 \beta k} - 1} +\int_{0}^{\infty} dk \frac{k^2}{E(k)} \frac{e^{\beta E(k)}}{e^{2 \beta E(k)} - 1} \right \}\\
&\equiv&\delta M_s^2.\nonumber
\ea
This implies that for all values of the bare mass and three-momentum, the scalar dispersion relation has the free form
\ba
\Erg = \sqrt{p^2 +M^2+\delta M_s^2},
\ea
where only the value of the mass shift $\delta M_s^2$ has any dependence on the temperature. If the bare mass $M$ is very large in comparison with $T$, the second integral in Eq.~(\ref{scalar}) becomes exponentially small and $\delta M_s^2 =  \frac{1}{2} g^2 C_F T^2$, while in the massless case, the two integrals contribute equally to yield a result twice as large. The full function $\delta M_s^2(M/T)$ interpolating between these two cases is difficult to obtain analytically, but is numerically readily available and is plotted in Fig.~\ref{fig1}.

\begin{FIGURE}[t]
{\centerline{\epsfxsize=10.0cm\epsfysize=6.5cm\epsfbox{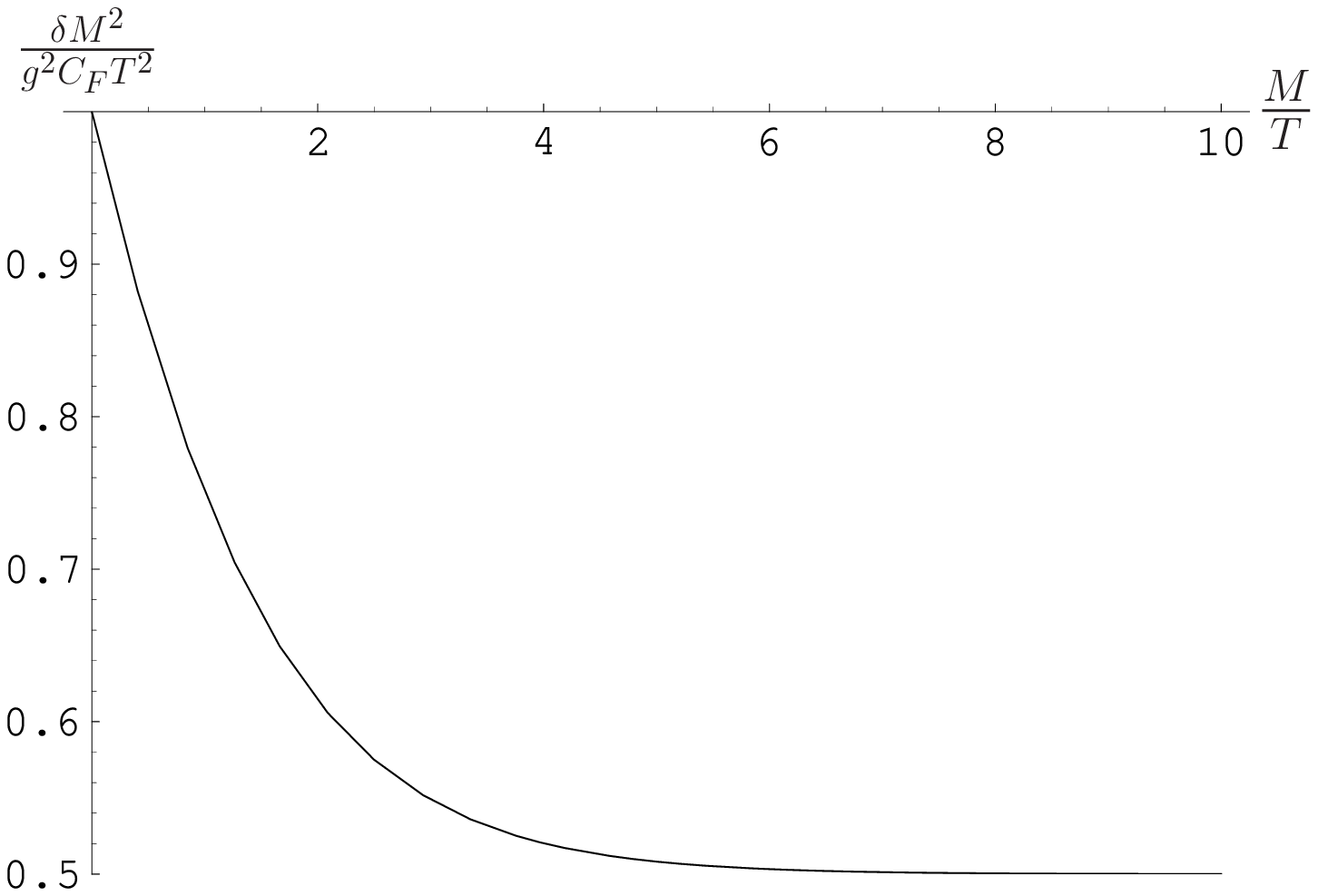}}

\caption[a]{The form of the leading order thermal correction to the mass squared of a fundamental ${\mathcal N}=2$ scalar field, $\delta M_s^2$, as defined in Eq.~(\ref{scalar}). We give the result as a function of $M/T$, normalized by $g^2C_F T^2$. \label{fig1}}}
\end{FIGURE}

\section{Fundamental quarks in ${\mathcal N} =4$ SYM \label{symq}}

Next, we inspect the case of fundamental ${\mathcal N} =2$ quarks in the SYM theory, and thus consider the leading order fermion self energy as given by the three graphs of Fig.~\ref{graphs1}.b. Continuing to work in the Feynman gauge, we observe that their sum can be written as
\ba
\label{sigma}
\Sigma(P) &=& 4 g^2 C_F \sintp \frac{i \left ( \slashed P + \slashed K \right ) +M}{(P+K)^2+M^2} \frac{1}{K^2} \nn
&-&4 g^2 C_F \sintm  \frac{i  \slashed K }{(P+K)^2+M^2} \frac{1}{K^2},
\ea
from which we obtain the equation for the pole of the propagator,
\ba
\label{pole1}
\det \left ( i \slashed P + M +\Sigma(P) \right ) &=& 0.
\ea

Looking at Eq.~(\ref{sigma}), we note that $\Sigma(P)$ may be written in the form
\ba
\Sigma(P) &=& \left ( i \slashed P + M \right ) a(P) + i \slashed V(P),
\ea
where the functions $a(P)$ and $V_\mu(P)$ are defined by
\ba
a(P)&=&4 g^2 C_F \sintp \frac{1}{(P+K)^2+M^2} \frac{1}{K^2},\\
V_\mu(P)&=& 4 g^2 C_F \bigg\{\sintp \frac{K_\mu}{(P+K)^2+M^2} \frac{1}{K^2}\nn
&-&\sintm  \frac{K_\mu}{(P+K)^2+M^2} \frac{1}{K^2}\bigg\}.
\ea
This implies that the fermion dispersion relation can be solved from the equation
\ba
\label{pole11}
\left ( P (1+a(P)) + V(P) \right )^2 + M^2 (1+a(P))^2 &=& 0,
\ea
which can be further simplified by noting that a simple power counting exercise shows that both $a(P)$ and $V_\mu(P)$ are regular and at most of order $g^2T/p_0$, when $P$ is on shell. The terms in the dispersion equation involving $a(P)$ are seen to be either proportional to $P^2+M^2$, and thus vanish at this order, or of the form $a\,P\cdot V$, which is always parametrically small. We conclude that we may thus neglect $a(P)$ altogether in our study and write Eq.~(\ref{pole11}) in the form
\ba
\label{pole}
\left ( P + V(P) \right )^2 + M^2  = 0,
\ea
which motivates us to next analyze the behavior of the functions $P\cdot V(P)$ and $V(P)^2$ in various kinematic regimes.

\subsection{The sum-integrals}

\subsubsection{$P\cdot V$}

To inspect the integral $P\cdot V$, we first separate the vector $V(P)$ into its bosonic and fermionic pieces by writing $V =V^{+} + V^{-}$, where the latter two functions read
\ba
V_{\mu}^{\pm}(P) = \pm 4 g^2 C_F \sintpm \frac{ K_{\mu}}{(P+K)^2 + M^2} \frac{1}{K^2}.
\ea
Completing squares and utilizing the results of Appendix \ref{sumints}, we obtain from here
\ba
&&P \cdot V^{\pm} = \\ \nonumber \\ \nonumber
&& \frac{g^2 C_F}{2\pi^2}\bigg\{ \real \int_{0}^{\infty} dk k \left (2 +\frac{P^2+M^2}{2 p k} \log \left ( \frac{ P^2 + M^2 +2 i k p_0 -2 pk}{P^2 + M^2 +2 i k p_0 +2 pk} \right ) \right ) \frac{1}{e^{\beta k} \mp 1} \\ \nonumber \\ \nonumber
&+& \real\int_{0}^{\infty} dk \frac{k^2}{E(k)} \left ( 2 +\frac{P^2 + M^2}{2 p k} \log \left ( \frac{ -P^2 +M^2 + 2 i E(k) p_0 -2 p k}{-P^2 +M^2 + 2 i E(k) p_0 +2 p k} \right ) \right )  \frac{1}{e^{\beta E(k)}  \pm 1}\bigg\}.
\ea
As discussed already in Section \ref{scalarsec}, upon taking the real parts of the logarithms, the integrals in the terms proportional to $P^2 + M^2$ are regular in the limit $P^2 \rightarrow -M^2$, and may consequently be ignored. This means that in the unperturbed on shell limit, we have to leading order in the coupling
\ba
\label{PV}
P \cdot V^{\pm} \big |_{P^2 + M^2 = 0} &=& \frac{g^2 C_F}{ \pi^2}\bigg\{ \int_{0}^{\infty} dk k\frac{1}{e^{\beta k} \mp 1} + \int_{0}^{\infty} dk \frac{k^2}{E(k)} \frac{1}{e^{\beta E(k)} \pm 1}\bigg\},
\ea
implying that however small the three-momentum $p$ becomes, $P \cdot V^{\pm}$ are always constants of order $g^2$.

\subsubsection{$V^2$}

In evaluating the $V(P)^2$ term in Eq.~(\ref{pole}), we note that we may from the beginning take both $p$ and $M$ to be of order $g T$ or higher, as for larger momenta and bare masses this term is certainly subdominant with respect to $P\cdot V$. Concentrating first on $V_0(P)$, we obtain
\ba
\label{V0} \nonumber
V_{0}^{\pm}(P) & = &   -\frac{g^2 C_F}{2 \pi^2p}\; \im \, \Bigg\{ \int_{0}^{\infty} dk k \left[ \log \left (\frac{ P^2 + M^2 +2 i k p_0 + 2 p k}{P^2 + M^2 +2 i k p_0 - 2 p k} \right ) \frac{1}{e^{\beta k} \mp 1}  \right. \\ \nonumber \\
&+& \left. \frac{E(k) + ip_0}{E(k)} \log \left ( \frac{ -P^2 + M^2 + 2 i E(k) p_0 + 2 p k}{-P^2 + M^2 + 2 i E(k) p_0 - 2 p k} \right )
\frac{1}{e^{\beta E(k)} \pm 1} \right] \Bigg\},
\ea
where $\im f(p_0) \equiv [f(p_0)-f(-p_0)]/(2i)$. We note that the integrals appearing here obtain their dominant contributions from the region $k\sim T$, implying that we may replace the logarithms in the unperturbed on shell limit by
\ba
\!\!\!\!\!\!\!\!&&\!\!\!\!\!\!\!\!\im\,\log \left ( \frac{ P^2 + M^2 +2 i k p_0 +2 pk}{P^2 + M^2 +2 i k p_0 -2 pk} \right ) = -i \log\left( \frac{ip_0 +p}{ip_0 -p}\right), \\ \!\!\!\! \nn
\!\!\!\!\!\!\!\!&&\!\!\!\!\!\!\!\!\im\,\Bigg[\frac{E(k) + ip_0}{E(k)}\log \left ( \frac{ -P^2 + M^2 + 2 i E(k) p_0 + 2 p k}{-P^2 + M^2 + 2 i E(k) p_0 - 2 p k} \right )\Bigg] \approx -i \log\left( \frac{ip_0 +p}{ip_0 -p}\right),
\ea
as well as set $M\rightarrow 0$ inside the thermal distribution functions.

Substituting now the above expressions to Eq.~(\ref{V0}), we obtain
\ba
\label{small}
V_{0}^{\pm}(P)&=&   \frac{ig^2 C_F}{2 \pi^2 p}\log \left (\frac{ip_0 +p}{ip_0 - p}\right )
\int_{0}^{\infty} dk k \left ( \frac{1}{e^{\beta k} \mp 1} + \frac{1}{e^{\beta k} \pm 1} \right ) \\ \nonumber \\ \nonumber
&=&\frac{i g^2 C_F T^2} {8p} \log \left (\frac{ip_0 +p}{ip_0-p} \right ),
\ea
from which the value of $\bm v^{\pm}(P)$ can be deduced as well. Rotational invariance namely implies that
\ba
\bm v^{\pm}(P) = \frac{\bm v^{\pm} \cdot \bm p}{p^2} \,\bm p=\frac{V^{\pm} \cdot P - V_0^{\pm} p_0}{p^2} \,\bm p,
\ea
where the $V^{\pm} \cdot P$ term in the numerator can be read off from the $M\ll k$ limit of Eq.~(\ref{PV}),
\ba
\label{PV2}
P \cdot V^{\pm} \big |_{P^2 + M^2 = 0} &=& \frac{1}{4} g^2 C_F T^2.
\ea
Putting everything together, we obtain to leading order
\ba
\label{vpm}
\bm v^{\pm} = \frac{1}{4 p^2} g^2 C_F T^2 \left (1 - \frac{ip_0}{2 p} \log \left ( \frac{ip_0+p}{ip_0-p} \right ) \right ) \bm p.
\ea
It is interesting to note that in the limit under consideration $V^{+} = V^{-}$, implying that both bosons and fermions contribute equally to the vector $V_\mu(P)$.

\subsection{Soft masses and momenta}

\begin{figure}[t]

\centerline{\epsfxsize=10.0cm\epsfysize=6.5cm\epsfbox{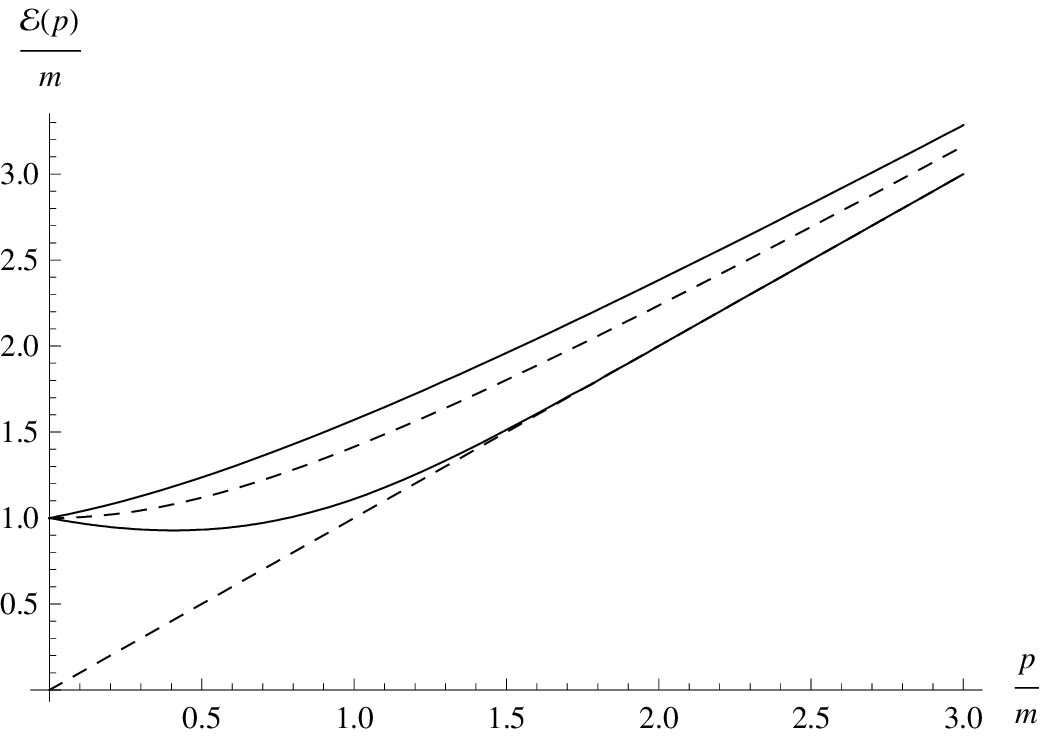}}

\caption[a]{Plots of the dispersion relation of a massless fermion in the SYM theory, $\Erg(p)$, as a function of $p/m$ with $m\equiv\sqrt{C_F/2}\,gT$. The solid lines correspond to the particle (upper curve) and plasmino (lower curve) modes, while the dashed lines show the free massive (upper curve) and massless (lower curve) dispersion relations, $\Erg(p)=\sqrt{p^2+m^2}$ and $\Erg(p)=p$, respectively. In the large $p$ limit, the particle excitation curve approaches the form $\Erg(p)=\sqrt{p^2+2m^2}$. \label{masslessquarks}}
\end{figure}

Let us begin the assembling of the dispersion relation from the case of massless fermions, $M=0$, and three-momenta of order $gT$. In this regime, the dispersion relation as solved from Eq.~(\ref{pole}) clearly becomes (with $p_0 \rightarrow i\Erg$)
\ba
\label{masslessdisp}
\Erg - i V_0 = \pm\(p+\frac{\bm v \cdot \bm p}{p}\),
\ea
from which we further obtain
\ba
m^2p +\(\Erg\pm p\)\bigg\{m^2\log\left ( \frac{\Erg-p}{\Erg+p} \right ) \pm 2p^2\bigg\} &=&0,\label{symqres}
\ea
with $m^2 = g^2 C_F T^2/2$. The two solutions of this equation, corresponding to the $+$ and $-$ signs above, are plotted in Fig.~\ref{masslessquarks}, where they are compared to the dispersion relations of free particles with masses $m$ and $0$.

The form of the dispersion equation (\ref{symqres}) is equivalent to what has been found for quarks in QCD previously \cite{weldon1}, except that the effective mass parameter $m$ appearing in the result is two times larger in the SYM theory. The two solutions corresponding to the minus and plus signs in the above equation are commonly dubbed as the so-called particle and plasmino modes, respectively. Of them, only the former one develops into a regular particle excitation at larger momenta, while the latter can be identified as a collective excitation that is negligible for $p\gg gT$. This can be verified by evaluating the residues of the two poles of the fermion propagator, which can be related to the creation probabilities of the respective excitations (see \textit{e.g.~}Ref.~\cite{petit}).

The case of fermionic excitations with non-zero, but soft masses $M\sim gT$ can on the other hand be studied by numerically solving Eq.~(\ref{pole}) for various values of the mass parameter. The results turn out to be highly similar to the QCD case, which has been investigated in Refs.~\cite{petit,pisa}, and we omit the details of the calculations here. The main conclusion drawn from them is that with increasing $M$, the form of the dispersion relation of the particle excitation very quickly approaches that of a free particle with a shifted mass,
\ba
\Erg &=&\sqrt{p^2+\widetilde{m}^2},\\
\widetilde{m}&\equiv&\fr{1}{2}\Big\{M+\sqrt{M^2+4m^2}\Big\},
\ea
and already for $M=m$, this approximation practically coincides with the numerical solution of Eq.~(\ref{pole}).

\subsection{Hard masses and momenta}

Consider now the case, where either the bare mass $M$ or the external three-momentum $p$ of the ${\mathcal N}=2$ quark is of order $T$. In this regime, $V^{\pm}$ clearly become of order $g^2$, so the terms in Eq.~(\ref{pole}) proportional to $V^2$ can be neglected to yield a simpler equation for the dispersion relation,
\ba
P^2 + 2 P \cdot V + M^2 = 0.
\ea
Using our earlier result for $P \cdot V^{\pm}$ from Eq.~(\ref{PV}), we see from here that the dispersion relation now becomes equal to that of the ${\mathcal N}=2$ scalars and reads
\ba
\Erg = \sqrt{p^2 +M^2+\delta M_q^2},
\ea
where
\ba
\delta M_q^2 &=&\delta M_s^2 \nn
&=& 2 g^2 C_F \left \{ \frac{1}{\pi^2} \int_{0}^{\infty} dk k \frac{2 e^{\beta k}}{e^{2 \beta k} - 1} + \frac{1}{\pi^2} \int_{0}^{\infty} dk k \frac{k}{E(k)} \frac{2 e^{\beta E(k)}}{e^{2 \beta E(k)} - 1} \right \}. \label{symqresx}
\ea
The behavior of this function has been studied already in Section \ref{scalarsec}.

\section{Quarks in QCD}

Next, we move on to study fundamental quarks in QCD. To leading order in the coupling, the quark self energy can be obtained by evaluating the single graph of Fig.~\ref{graphs1}.c. Continuing with the Feynman gauge, we obtain for this function
\ba
\Sigma(P) &=& 2 g^2 C_F \sintp \frac{i (\slashed P + \slashed K) + 2 M}{(P+K)^2 + M^2} \frac{1}{K^2} \label{qcdself1} \, ,
\ea
which can clearly be written in the form  $\Sigma(P)   = \big(i\slashed V^+(P) + ( i \slashed P + 2 M ) a(P)\big)/2$, where $V^+(P)$ and $a(P)$ stand for the functions introduced in the previous Section. The corresponding dispersion relation is then given by the solution to
\ba
\label{qcddis}
(P(1+ a(P)/2)+V^+(P)/2)^2 + M^2(1+a(P))^2 = 0,
\ea
which we now proceed to investigate in various limits. The case of soft masses and momenta has already been worked out in the literature in Refs.~\cite{weldon1,petit,pisa}, but we reproduce it below as well for completeness.

\subsection{Soft masses and momenta}

Let us start from the case of unbroken chiral invariance, $M=0$, and three-momenta of order $gT$. In this limit, the calculation reduces to that presented in the previous Section, as we note that we may again altogether neglect the function $a(P)$. The dispersion relation is then given by the solutions to the equation
\ba
\label{qcdmasslessdisp}
\Erg - i V_0^+/2 = \pm\(p+\frac{\bm v^+ \cdot \bm p}{2p}\),
\ea
from which we again solve
\ba
2\widetilde{m}^2p +\(\Erg\pm p\)\bigg\{\widetilde{m}^2\log \left ( \frac{\Erg-p}{\Erg+p} \right ) \pm 2p^2\bigg\} &=&0. \label{qcdres1}
\ea
This time the mass parameter $\widetilde{m}$ is, however, smaller by a factor of $2$ in comparison with the SYM case,
\ba
\widetilde{m}^2 &=& g^2 C_F T^2/8.
\ea

The interpretation of the two modes corresponding to the $+$ and $-$ signs above is equivalent to the SYM case, and the behavior of $\Erg(p)$ is trivial to infer from Fig.~\ref{masslessquarks}. The case of non-zero but soft masses has been studied in Ref.~\cite{petit}, with the result that the collective plasmino mode quickly becomes negligible when the rest mass parameter exceeds the value $\widetilde{m}$.

\subsection{Hard masses and momenta}

Next, let again either the bare mass or the external three-momentum be of order $T$, which to the best of our knowledge has not been studied in the literature before. In this limit, the functions $a(P)$ and $V^+(P)$ are both regular and of order $g^2$ in the unperturbed on shell limit, so we have from Eq.~(\ref{qcddis}) the dispersion equation
\ba
\Erg^2&=&E(p)^2 + \Big[P \cdot V^+(P) +M^2a(P)\Big]_{P^2=-M^2}\, . \label{qcdres2}
\ea
Using the results presented in Section \ref{symq}, we immediately obtain for the first integral
\ba
P \cdot V^+(P)\big |_{P^2 + M^2 = 0} = \frac{g^2 C_F}{\pi^2} \int_{0}^{\infty} dk k\left \{  \frac{1}{e^{\beta k} - 1} +  \frac{k}{E(k)} \frac{1}{e^{\beta E(k)} +1} \right \},\label{pvqcd}
\ea
a result obviously independent of the three-momentum. Similarly, we get for the function $a(P)$
\ba
a(P) &=&   \frac{g^2 C_F}{2 \pi^2}\bigg\{  \real \int_{0}^{\infty}  dk \frac{1}{p}\log \left (\frac{P^2 + M^2 + 2 i k p_0 + 2 p k}{ P^2 + M^2 + 2 i k p_0 - 2 p k} \right ) \frac{1}{e^{\beta k} - 1} \\ \nonumber \\ \nonumber
&+& \real \int _{0}^{\infty} dk \frac{k}{p E(k)} \log \left ( \frac{-P^2 + M^2 +2 i E(k) p_0 + 2 p k}{-P^2 + M^2 +2 i E(k) p_0 - 2 p k} \right )  \frac{1}{e^{\beta E(k)} + 1}\bigg\},
\ea
in which the first logarithm vanishes upon taking its real part and then going to the on shell limit. From the second logarithm, we on the other hand get
\ba
a(P)\big |_{P^2 + M^2 = 0}&=&\frac{g^2 C_F}{4 \pi^2} \int _{0}^{\infty} dk \frac{k}{p E(k)} \log \left (\frac{k-p}{k+p} \right )^2 \frac{1}{e^{\beta E(k)} + 1}, \label{aint}
\ea
where we have used the result of Eq.~(\ref{logid}).

\begin{figure}[t]

\centerline{\epsfxsize=10.0cm\epsfysize=6.5cm\epsfbox{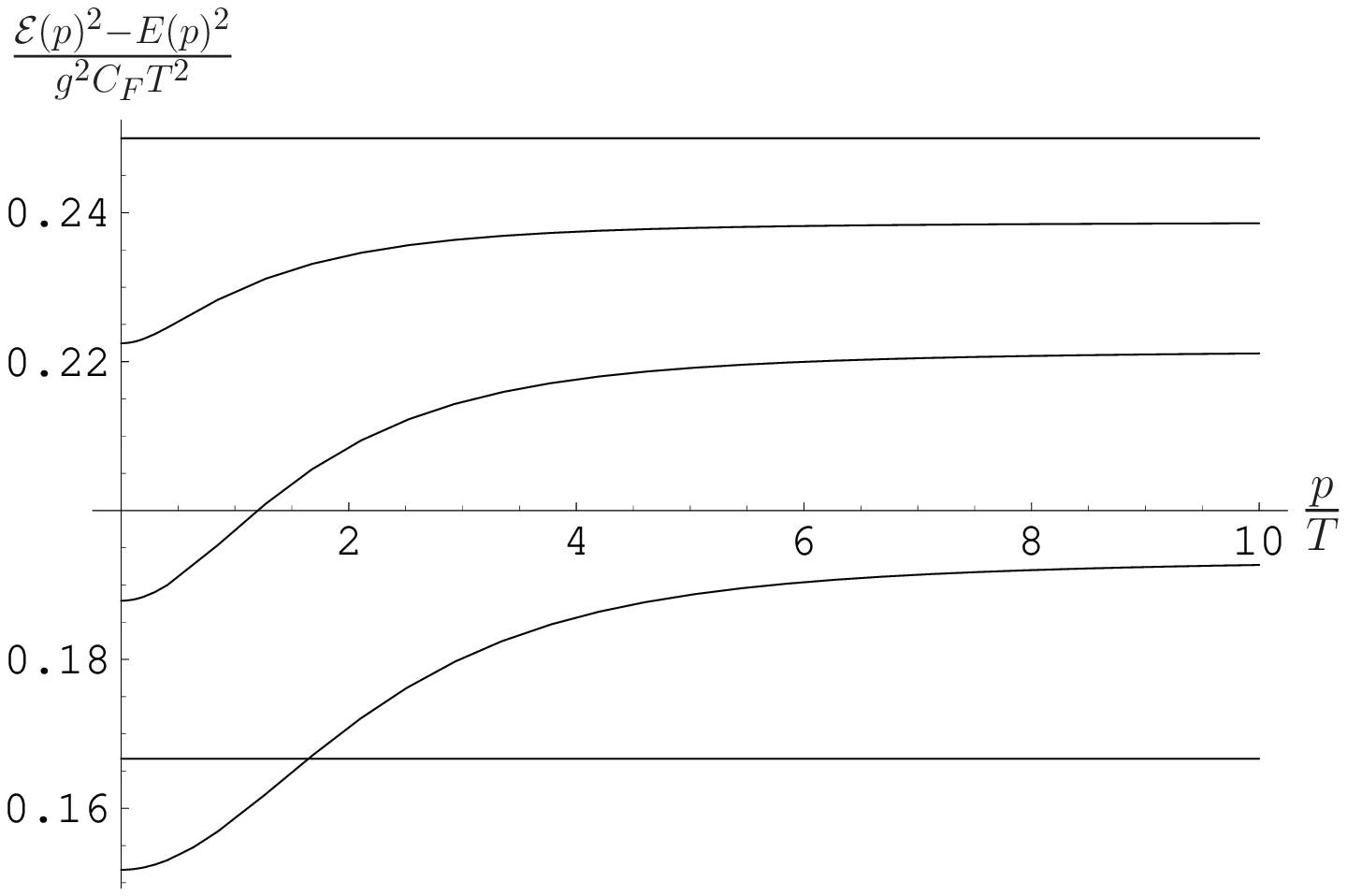}}

\caption[a]{The thermal correction to the non-interacting dispersion relation of a quark in QCD, $\Erg(p)^2 - E(p)^2$, as a function of $p/T$ for various values of $M/T$. Counting from top down on the right side of the plot, the curves correspond to $M/T=0,\; 1/2,\; 1,\; 2,\; \infty$. \label{qcdfig}
}
\end{figure}

First consider the case of $M=0$ but $p\sim T \gg gT$.  In this limit, $M^2 a(P)$ naturally vanishes, while $P \cdot V^+(P) = \frac{1}{4} g^2 C_F T^2$, so the dispersion relation reads
\ba
\Erg(p)^2 &=& p^2 + m^2,
\ea
with $m^2 =  \frac{1}{4} g^2 C_F T^2$.  Similarly, in the case with $M \gg T$, $a(P)$ as well as the second term in Eq.~(\ref{pvqcd}) become exponentially small, so the dispersion relation becomes
\ba
\Erg(p)^2 &=& p^2 +M^2+\delta M^2, \label{highm1}
\ea
with
\ba
\delta M^2 &=&  \frac{1}{6} g^2 C_F T^2.
\ea
Between these two limits, we must resort to a numerical evaluation of the functions $P \cdot V^+(P)$ and $a(P)$, and obtain the dispersion relation $\Erg(p)$ displayed in Fig.~\ref{qcdfig}.

Finally, we note for completeness that if $M$ is of order $T$, but $p \sim gT$, we may obviously expand Eq.~(\ref{aint}) in the limit of small $p/k$. Using Eq.~(\ref{qcdres2}), this leads to the dispersion relation
\ba
\Erg(p)^2 &=& p^2 +M^2+\delta M^2,\\
\delta M^2&=&\frac{g^2 C_F}{\pi^2} \int_{0}^{\infty} dk \left \{  \frac{k}{e^{\beta k} - 1} +  \frac{k^2-M^2}{E(k)} \frac{1}{e^{\beta E(k)} +1} \right \},
\ea
where $\delta M^2$ corresponds to the $p\rightarrow 0$ limit of the functions displayed in Fig.~\ref{qcdfig}.

\section{The non-relativistic heavy particle limit \label{nonrellimit}}

As we have seen above, the leading $\mathcal O(g^2)$ thermal corrections to the dispersion relations of heavy fundamental quarks and scalars are suppressed by an inverse power of the bare mass and consequently vanish as $M \rightarrow \infty$. At the same time, on physical grounds one would expect the thermal corrections coming from soft loop momenta of order $gT$ to survive. To see this, consider the interaction of a heavy charged particle with an electromagnetic field, and in particular, how thermal effects modify the long distance behavior of the fields sourced by the particle. Debye screening cuts the longitudinal electric fields off at a distance of order the inverse Debye mass $m_D$, which can easily be seen to give the energy stored in the electric fields a temperature dependent contribution of order $g^2 m_{D}$. The factor $g^2$ here comes from the fact that the electric fields sourced by the quark are $\mathcal O(g)$, so the electromagnetic energy density (which depends on the square of the electric fields) is $\mathcal O(g^2)$, while the factor $m_D$ is a direct consequence of the finite range $1/m_D$ of the static fields.\footnote{Due to the Debye screening, the energy stored in the fields becomes proportional to an integral of the form $\int {\rm d}^3p\,\fr{1}{p^2+m_D^2}$.} We therefore see that at cubic order in the coupling, one should find thermal corrections to the heavy particles' energy (and thus their dispersion relation), which are independent of the heavy mass and in fact become the dominant contributions for bare masses of order $T/g$ and larger.

\begin{figure}[t]

\centerline{\epsfxsize=14.0cm\epsfbox{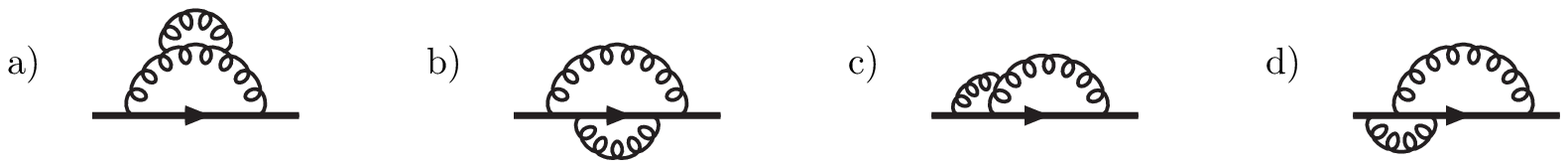}}

\caption[a]{
Four different Feynman diagrams depicting possible two-loop corrections to the heavy quark self energy in QCD. As is explained in the text, only the graph a is observed to lead to an ${\mathcal O}(g^3)$ contribution to the quantity.
\label{dresseddiagrams}
}
\end{figure}

To put the above statements on a more quantitative footing, consider now the QCD type diagram of Fig.~\ref{graphs1}.c, which stands for a generic example of the one loop graphs contributing to the heavy fermion/scalar self energies in both QCD and SYM. The IR sensitive region of momentum space that might upon a suitable resummation lead to the suspected ${\mathcal O}(g^3)$ behavior of the quantity is one where the gluon line is soft, \textit{i.e.~}$q_0\sim q\sim gT$. To inspect which types of loop corrections the graph may be sensitive to, we add an extra gluon line to dress either the light field propagator, the heavy field propagator, or the vertex, as displayed in Fig.~\ref{dresseddiagrams}, and investigate, what power of $g$ these corrections are suppressed by. It is easy to convince oneself that the qualitative results obtained here will also apply to the other diagrams of Fig.~\ref{graphs1}.a--b with the same topology, where either a gluon or a light scalar line is dressing the heavy quark or scalar. A brief inspection on the other hand shows that the graphs involving either a light fermion propagator or a four boson coupling cannot contribute to the self energy at the order we are interested in.


For the graph of Fig.~\ref{dresseddiagrams}.a, which is a generic example of diagrams containing loop corrections to the gluon propagator in Fig.~\ref{graphs1}.c, we observe that the maximal contributions originate, when the four-momentum of the added propagator is hard, $K\sim T$. Comparing this to the original one-loop graph, we observe that it obtains an extra factor of $g^2$ from the two new vertices, as well as a factor $1/g^2$ from the induced soft gluon propagator with momentum $\sim gT$ (recall that we assumed the momentum flowing along the gluon line of the original graph to be soft). We therefore conclude that the addition of the extra gluon line contributes at relative order $g^0$ and that loop corrections to the soft gluon line of the graph of Fig.~\ref{graphs1}.c must thus be resummed. This resummation may be carried out using the hard thermal loop (HTL) approximation for the gluon self energy, as the momentum flowing along the gauge boson line of the original one-loop diagram is soft, while the momentum flowing in the loops of the self energy is hard.

\begin{figure}[t]

\centerline{\epsfxsize=14.0cm\epsfbox{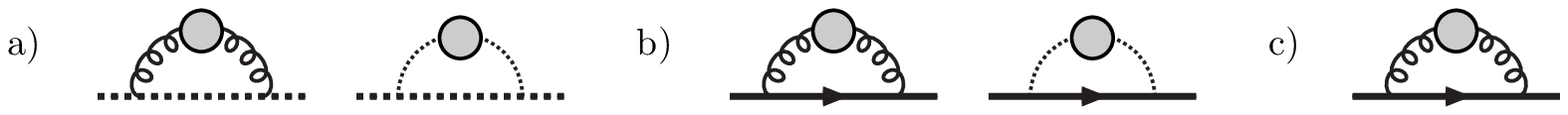}}

\caption[a]{
The diagrams contributing to the self energy of an infinitely massive fundamental a) scalar and b) fermion in the ${\mathcal N}=4$ SYM theory, and c) a quark in QCD. The intermediate light propagators are dressed with the HTL self energy function, which for the light scalars implies the use of the scalar thermal mass $m_s$. In the limit of a large quark mass $M\gtrsim T/g$, the graphs are dominated by soft light quark momentum $q_0\sim q \sim gT$. \label{dressedpropdiagram}}
\end{figure}

For the graphs b--d, it is straightforward to see that similar order-of-magnitude estimates, both for hard and soft loop momenta, always lead to the suppression of the result by at least one relative power of $g$. This behavior can be seen to carry on to further perturbative orders as well, thus confirming that it will not be necessary to carry out resummations amounting to dressing the heavy field propagator or the vertex function in the original diagram. We thereby see that in order to obtain the ${\mathcal O}(g^3)$ contributions to the self energy of a heavy quark in QCD, we only need to consider the graph of Fig.~\ref{dressedpropdiagram}.c, where the gluon line has been dressed with the HTL self energy. This diagram, as well as the corresponding ones in Figs.~\ref{dressedpropdiagram}.a and b, needed in the SYM computation, will be evaluated below in Section~\ref{QFTcalculation}. In the graphs containing a light scalar exchange, the HTL self energy reduces to a scalar thermal mass squared, $m_s^2$.


Before proceeding to the resummation calculation, we note that the above reasoning also implies that it is possible to determine the thermal corrections to the dispersion relation of a heavy particle using classical field theory. As the momentum flowing along the dressed boson lines in Fig.~\ref{dressedpropdiagram} is parametrically smaller than $T$, thermal fluctuations of the light fields sourced by the heavy particle should be negligible and the dynamics of these fields governed by classical field theory. The necessity to dress the light propagators with the HTL self energy, however, implies that the fields will experience thermal effects --- the classical equations of motion and the classical Hamiltonian will depend on thermal susceptibilities corresponding to the HTL self energy function. For a gauge field, the classical equations of motion will be those of macroscopic electrodynamics with temperature dependent permittivity and permeability tensors, which can be straightforwardly obtained from the HTL self energy function. For a scalar field, the equation of motion will on the other hand simply be the Klein-Gordon equation with a temperature dependent screening mass, which is nothing but the zero momentum limit of the light scalar field self energy function.

We will now proceed to consider the classical calculation outlined above, with the objective being the derivation of the correct large-$M$ limit of the dispersion relation of the fundamental particle. After this, we will in Section~\ref{QFTcalculation} verify the result by performing the corresponding diagrammatic computation, explicitly evaluating the graphs of Fig.~\ref{dressedpropdiagram}. As we are here considering heavy particles with typical momenta $p\sim T\ll M$, we will in both cases work in the non-relativistic limit and thus restrict ourselves to determining the thermal shifts to the rest and kinetic masses of the excitation.

\subsection{Classical field theory}

With the above arguments in mind, consider now a large volume $V$, which encloses a heavy charged particle moving along some trajectory. The energy contained in $V$ will depend on how fast the particle is moving. As we are taking $M \rightarrow \infty$, it is convenient to consider the dispersion relation as a function of the particle's velocity $v$ rather than its momentum. Classically, the energy contained in $V$ is
\ba
\mathcal E(v) &=& \frac{M}{\sqrt{1 - v^2 }} + \Delta\mathcal E_{\rm field}(v), \label{61}
\ea
where $\Delta\mathcal E_{\rm field}$ is the shift in the energy stored in the light scalar and/or electromagnetic fields sourced by the charge due to screening effects.\footnote{The renormalized mass $M$ already contains the energy of the fields in the vacuum.} By rotational invariance, the field energy must have the small $v$ expansion
\ba
\label{classicalenergy}
\Delta\mathcal E_{\rm field}(v) &=& \delta M_{\rm rest} + \frac{1}{2}\delta M_{\rm kin} v^2 + \mathcal O(v^4).
\ea
This equation, which is manifestly finite in the large $M$ limit, {\em defines} the shift in the rest and kinetic mass of the particle.

In what follows, we will compute the contributions to the rest and kinetic masses of the heavy fundamental particle from the light scalar and electromagnetic sectors separately, and then assemble the results for quarks in QCD as well as quarks and scalars in SYM. We will henceforth refer to the heavy particle as a quark, but this is no restriction, as the classical calculation is not sensitive to the spin of the heavy particle.

\subsubsection{Scalar field contributions}

Let $\bm x_{\rm quark}(t)$ now denote the trajectory of a heavy quark with mass $M$. In theories such as ${\mathcal N=4}$ SYM, the quark's coupling to the light scalar fields is dictated by an effective Lagrangian of the type
\ba
\label{classicallag}
L_{\rm classical} &=& -M \sqrt{1-\dot x_{\rm quark}^2 } + \int d^3 x
\left [  \frac{1}{2} \left (\dot \phi^2 - (\nabla \phi)^2 - m_s^2 \phi^2 \right ) - \rho \phi \right ],
\ea
where
\ba
\rho &=& e_{\rm eff} \sqrt{1-\dot x_{\rm quark}^2} \;\delta^3(\bm x - \bm x_{\rm quark}(t))
\ea
is the charge density of the quark, which must be a Lorentz scalar. The constant $e_{\rm eff}$ denotes the effective coupling of the fundamental particles to the light scalar fields, and through matching to the full theory\footnote{An easy way to see that this is the correct value of $e_{\rm eff}$ is to compute the energy density stored in scalar fields at distances $d \ll 1/T$ from the quark and match the resulting expression to the full theory result.} is easily seen to read $e_{\rm eff}^2 = C_F g^2$. The parameter $m_s$ on the other hand corresponds to the thermal screening mass of the light scalars, and is thus a function of the temperature. Its value is given by
\ba
m_s^2 = \lim_{Q \rightarrow 0} \Pi_{\rm s}(Q),
\ea
where $\Pi_{\rm s}(Q)$ is the light scalar self energy function and the limit of vanishing four-momentum is independent of the order of taking $p_0$ and $p$ to zero.

From the Lagrangian of Eq.~(\ref{classicallag}), we easily obtain the classical Hamiltonian
\ba
H_{\rm classical} &=& \frac{M}{\sqrt{1 -\dot x_{\rm quark}^2 }} + \mathcal E_{\rm scalar},
\ea
where
\ba
\label{classicalscalarenergy}
\mathcal E_{\rm scalar} &=&  \int d^3 x \left [ \frac{1}{2} \left ( \dot \phi^2 + (\nabla \phi)^2 + m_s^2 \phi^2\right )  - \frac{1}{1 -\dot x_{\rm quark}^2} \rho \phi \right ],
\ea
is the energy stored in the scalar field. This leads to a sourced Klein-Gordon equation of motion for the field,
\ba
\left (-\partial^2 + m_s^2 \right ) \phi = \rho,
\ea
which, taking the quark's velocity $\bm v \equiv \dot {\bm x}_{\rm quark}$ to be constant and introducing a spatial Fourier transform, has the solution
\ba
\phi(t,\bm k) &=& e_{\rm eff} \sqrt{1- v^2} \frac{e^{-i \omega_0(\bm k) t}}{-\omega_0(\bm k)^2 + k^2 + m_s^2},
\ea
with $\omega_0(\bm k) \equiv \bm v \cdot \bm k$. Substituting now this into Eq.~(\ref{classicalscalarenergy}), we see that we can finally write the classical field energy in the simple form
\ba
\label{classicalscalarenergy2}
\mathcal E_{\rm scalar} = \frac{e_{\rm eff}^2}{2} \int \frac{d ^3 k}{(2 \pi)^3}
\frac{2 \omega_0^2 -v^2 \left (\omega_0^2 + k^2 + m_s^2 \right ) }{\left (- \omega_0^2 +\bm k^2 + m_s^2 \right )^2},
\ea
which we however immediately observe to be UV divergent.

To cure the divergence, we note that it is independent of the screening mass $m_s$ and hence the temperature $T$, and corresponds to the singular $\sim1/x^4$ energy density of a localized point particle. We may thus simply subtract from the field energy the corresponding expression with $m_s = 0$,
\ba
\Delta \mathcal E_{\rm scalar} \equiv  \mathcal E_{\rm scalar}- \mathcal E_{\rm scalar} \big |_{m_s = 0}\, ,
\ea
which is allowed as we are only interested in the shift in the field energy due to the presence of screening. The resulting three-dimensional momentum integral is both finite and straightforward to evaluate, and leads to the simple result
\ba
\Delta \mathcal E_{\rm scalar} = \frac{e_{\rm eff}^2 m_s}{8 \pi} \frac{1}{\sqrt{1-v^2}}.
\ea
Taking the small $v$ limit here and comparing to Eq.~(\ref{classicalenergy}), we see that the shift in both the rest and kinetic masses of the heavy quark due to its interaction with the light scalar fields is equal to
\ba
\label{classicalscalar}
\delta M &=& \frac{e_{\rm eff}^2 m_s}{8 \pi}.
\ea
This equality is a consequence of the Lorentz invariant nature of the influence of $\Pi_{\rm light}$ on the classical equations of motion, as in the long wavelength limit it reduces to the Lorentz preserving screening mass $m_s$. As we will discuss below, this is however not the case for electromagnetic fields.

\subsubsection{Electromagnetic field contributions}

The derivation of the energy contained in electromagnetic fields is slightly more involved than it was for the case of scalar fields. The reason is that even in the limit of a soft external momentum, the gluon polarization tensor has nontrivial momentum dependence, which results in complex permittivity and permeability tensors, and correspondingly non-trivial dispersion and damping of the electromagnetic fields. The presence of dispersion and damping makes the construction of a Hamiltonian from the Lagrangian formulation of the electromagnetic problem impossible. However, as we elaborate on below, using the classical equations of motion and the expected form of the energy conservation equation in the presence of dissipation, one can extract a meaningful definition of the electromagnetic energy of a moving quark.\footnote{The argument we use below is a straightforward generalization of that found in Refs.~\cite{Jackson,LL}. Similar issues have also been discussed in Ref.~\cite{Blaizot:1994am}.}

We begin the calculation by noting that in the limit of weak fields, the non-linear non-Abelian interaction terms appearing in the classical Yang-Mills equations of motion may be neglected. This leaves as the linearized equations of motion of the electric and magnetic fields the macroscopic Maxwell's equations in a medium,
\begin{subequations}
\label{maxwell}
\begin{align}
\nabla \cdot \bm D &= \rho, \  &  \nabla \cdot \bm B &= 0,
\\
\nabla \times \bm E &= - \frac{\partial \bm B}{\partial t},  \  & \nabla \times \bm H &= \bm J + \frac{\partial \bm D}{\partial t},
\end{align}
\end{subequations}
where the macroscopic fields $\bm D$ and $\bm H$ are defined in Fourier space ($k_0\equiv \omega$ here) via
\begin{eqnarray}
\bm D(\omega,\bm k) &=&\bm  \epsilon(\omega,\bm k) \bm E(\omega,\bm k),
\\
\bm B(\omega,\bm k) &=& \bm \mu(\omega,\bm k) \bm H(\omega,\bm k),
\end{eqnarray}
and $\bm \epsilon$ and $\bm \mu$ are the permittivity and permeability tensors describing the medium. We note that even in the limit of small momentum, $\bm \epsilon(\omega,\bm k)$ and $\bm \mu(\omega,\bm k)$ still have nontrivial momentum dependence, as they become functions of $\omega/k$. The source $J^{\mu} = (\rho, \bm J)$ is on the other hand simply the current of a massive localized quark,
\begin{eqnarray}
\rho &=& e_{\rm eff} \delta^{3}(\bm x - \bm x_{\rm quark}(t)),
\\
\bm J &=& e_{\rm eff} \frac{d \bm x_{\rm quark} }{dt }\delta^{3}(\bm x - \bm x_{\rm quark}(t)).
\end{eqnarray}

Consider now a large volume $V$, which encloses the source $J^{\mu}$ and whose size is larger than any characteristic length scale associated with the medium. In this volume, the energy conservation equation reads
\ba
\label{cons0}
\frac{d }{d t} \int_V d^3 x \, u_{\rm EM} &=& -\int_{V} d ^3 x \left \{ \nabla \cdot \bm S +  \bm J \cdot \bm E  \right \} - Q_{\rm heat},
\ea
where $u_{\rm EM}$ is the electromagnetic energy density, $\bm S = \bm E \times \bm H$ the Poynting vector, and $Q_{\rm heat}$ the energy lost to heat per unit time. This equation simply states that the rate at which the electromagnetic energy contained in $V$ decreases is given by the rate at which electromagnetic energy flows out of $V$ plus the work done by the source and the energy lost to heat. Using Maxwell's equations, we can furthermore express the divergence of the Poynting vector in the form
\ba
\nabla \cdot \bm S &=& - \bm E \cdot \frac{\partial \bm D}{\partial t} - \bm H \cdot \frac{\partial \bm B}{\partial t} - \bm J \cdot \bm E,
\ea
finally giving us
\ba
\label{energycons1}
\frac{d }{d t} \int_V d^3 x \, u_{\rm EM} &=& \int_{V} d ^3 x \left \{ \bm E \cdot \frac{\partial \bm D}{\partial t}
+  \bm H \cdot \frac{\partial \bm B}{\partial t}  \right \} - Q_{\rm heat}.
\ea

In a medium where there is no dispersion or dissipation ({\em i.e.~}where $\bm \epsilon$ and $\bm \mu$ are constant), the time derivatives appearing on the right hand side of Eq.~(\ref{energycons1}) can be moved out of the integral. In this case, we furthermore have $Q_\mathrm{heat} = 0$, and thus $u_{\rm EM} = \frac{1}{2}  \left (\bm E \cdot \bm D + \bm B \cdot \bm H \right ).$ If $\bm \epsilon$ and $\bm \mu$ are not constant, as is the case in the theories we are considering, the time derivative, however, cannot be factored out of the integrand, and more care must be taken in order to obtain the correct expression for the electromagnetic energy.

To this end, let us now introduce a spacetime Fourier transform and assume that the electromagnetic fields may be written in the form
\begin{align}
\bm E(\omega,\bm k) &= \widetilde {\bm E}(\omega,\bm k) f(\omega-\omega_0(\bm k)),
\\
\bm H(\omega,\bm k) &=   \widetilde{\bm H}(\omega,\bm k) f(\omega-\omega_0(\bm k)),
\end{align}
where the real valued function $f(x)$, normalized such that $\int_{-\infty}^{\infty} {\rm d}x\, f(x) = 1$, has support only on a narrow interval around $x = 0$, and $\widetilde {\bm E}(\omega,\bm k)$ and $\widetilde {\bm H}(\omega,\bm k)$ are slowly varying functions around $\omega = \omega_0(\bm k)$.\footnote{Such an ansatz is certainly valid for fields sourced by a quark moving at a constant velocity $\bm v$, in which case they become proportional to $\delta (\omega - \bm v \cdot \bm k)$.} Under this assumption, we obtain
\begin{align}
 \label{step1}
&
\int_V d^3 x \, \big  \{ \bm E \cdot \frac{\partial \bm D}{\partial t} +\bm H \cdot \frac{\partial \bm B}{\partial t} \big \} =
 \\ \nonumber
&
 \int \frac{d \omega}{2 \pi} \frac{d \omega' }{2 \pi} \frac{d^3 k}{(2 \pi)^3}
\left  \{   -i \omega \widetilde {\bm E}^\dagger(\omega',\bm k)  \bm \epsilon(\omega, \bm  k)\widetilde {\bm E}(\omega,\bm k)
\right  \} f(\omega'-\omega_0(\bm k)) f(\omega-\omega_0(\bm k)) e^{i (\omega' {-} \omega)t}
\\ \nonumber
+&\int \frac{d \omega}{2 \pi} \frac{d \omega' }{2 \pi} \frac{d^3 k}{(2 \pi)^3}
\left  \{   -i \omega \widetilde {\bm H^\dagger}  (\omega',\bm k)  \bm \mu(\omega, \bm  k) \widetilde {\bm H} (\omega,\bm k)
\right  \}  f(\omega'-\omega_0(\bm k)) f(\omega-\omega_0(\bm k)) e^{i (\omega' {-} \omega)t},
\end{align}
where the frequency integrations will be dominated by $\omega \approx \omega' \approx \omega_0(\bm k)$. Expanding now the slowly varying part of the integrand in powers of $(\omega - \omega')$ --- which can be represented as a time derivative --- and integrating over the frequency variables, we find that at leading order the result reads
 \begin{align}
 \label{step2}
\int_V d^3 x \, \bigg \{ \bm E \cdot \frac{\partial \bm D}{\partial t} &+\bm H \cdot \frac{\partial \bm B}{\partial t} \bigg \}
 = \frac{d \mathcal E_{\rm EM} }{dt}  + Q_{\rm heat},
\end{align}
where we have defined
\begin{align}
\label{fieldenergy}
\nonumber
\mathcal E_{\rm EM} \equiv \frac{1}{8\pi^2}  \int & \frac{d^3 k}{(2 \pi)^3}   \bigg \{
\widetilde {\bm E}^{\dagger}(\omega,\bm k)   \partial_\omega \left [ \omega \bm \epsilon(\omega,\bm k)   \right ] \widetilde {\bm E}(\omega,\bm k)
+ \widetilde {\bm H}^{\dagger}(\omega,\bm k)  \partial_\omega \left [ \omega \bm \mu(\omega,\bm k)  \right ] \widetilde {\bm H}(\omega,\bm k)
\\
 &+ \omega \widetilde {\bm E}^{\dagger}(\omega,\bm k)  \bm \epsilon(\omega,\bm k)  \, \partial_\omega \widetilde {\bm E}(\omega,\bm k)
 - \omega \partial_\omega \widetilde {\bm E}^{\dagger} (\omega,\bm k)  \bm \epsilon(\omega,\bm k)  \widetilde {\bm E}(\omega,\bm k)
 \\
  &+ \omega \widetilde {\bm H}^{\dagger}(\omega,\bm k)  \bm \mu(\omega,\bm k)  \, \partial_\omega \widetilde {\bm H}(\omega,\bm k)
 - \omega \partial_\omega \widetilde {\bm H}^{\dagger}(\omega,\bm k) \bm \mu(\omega,\bm k)  \widetilde {\bm H} (\omega,\bm k)
 \bigg \} \bigg |_{\omega = \omega_0(\bm k)}, \nonumber
\end{align}
and
\ba
Q_{\rm heat} \,\equiv\, \fr{1}{4\pi^2}\int \frac{d^3 k}{(2 \pi)^3} \bigg \{&-&i  \omega
\widetilde {\bm E}^{\dagger}(\omega,\bm k)  \, \bm \epsilon(\omega,\bm k) \widetilde {\bm E}(\omega,\bm k)\nn
&-&i \omega \widetilde {\bm H}^{\dagger}(\omega,\bm k) \bm \mu(\omega,\bm k) \widetilde {\bm H}(\omega,\bm k)
 \bigg \} \bigg |_{\omega = \omega_0(\bm k)}.
\ea

Comparing the above expressions to the anticipated form of the energy conservation equation (\ref{energycons1}), we identify the quantity $\mathcal E_{\rm EM}$ with the electromagnetic field energy contained in $V$. In the limit, where $\bm \epsilon$ and $\mu$ are constant, only the first two terms in Eq.~(\ref{fieldenergy}) contribute to the integration, and as expected, we obtain $\mathcal E_{\rm EM} = \frac{1}{2} \int d^3 x \left (\bm E \cdot \bm D + \bm B \cdot \bm H \right )$. The quantity $Q_{\rm heat}$, which only depends on the imaginary parts of $\bm \epsilon$ and $\bm \mu$, is on the other hand clearly the energy lost to heat in the volume $V$ per unit time.

To obtain the shifts in the rest and kinetic masses of the quark, we now again specialize to the case, where its motion non-relativistic. In this limit, the magnetic field and the transverse component of the electric field created by the quark are $\mathcal O(v)$ already at $T = 0$, while thermal contributions to $\bm \mu$ and the transverse component of $\bm \epsilon$ are $\mathcal O(\omega^2)$ in the small momentum limit. We therefore conclude that any thermal effects in the electromagnetic field energy coming from magnetic and transverse electric fields are $\mathcal O(v^4)$, and we may concentrate on the contribution to the field energy coming from longitudinal electric fields.

Next, we solve for the longitudinal component of the electric field from Maxwell's equations by first introducing a spacetime Fourier transform. This leads us to
\ba
\label{fieldsol}
{\bm E}_L(\omega,\bm k)  &=& 2\pi e_\rmi{eff} \,\frac{i \bm k}{k^2 \epsilon_{L}(\omega, k) } \delta(\omega-\omega_0(\bm k)),
\ea
where again $\omega_0(\bm k) \equiv \bm v \cdot \bm k$, and $\epsilon_{L}$ is the longitudinal part of the permittivity tensor, which can be given in terms of the gluon polarization tensor as
\ba
\label{epsilonL}
\epsilon_{L}(\omega,\bm k) = 1 + \frac{\Pi_{00}(\omega,k)}{k^2}.
\ea
In the HTL limit, we on the other hand obtain for $\Pi_{00}$ \cite{Weldon:1982aq,Kalashnikov:1979cy}
\ba
\label{PI00}
\Pi_{00}(\omega,k) = m_{D}^2 \left (1 - \frac{\omega}{2 k} \left [ \log \left ( \frac{k + \omega}{k - \omega} \right )- i \pi \right ] \right ),
\ea
where $m_D$ is the Debye mass of the theory.

Putting finally Eq.~(\ref{fieldenergy}) and Eqs.~(\ref{fieldsol})--(\ref{PI00}) together and expanding the resulting expression in powers of $v$, we find that to second order in the velocity, the electromagnetic field energy is given by
\ba
\mathcal E_{\rm EM} = \frac{e_{\rm eff}^2}{2} \int \frac{d^3 k}{(2 \pi)^3} \left [\frac{1}{k^2 + m_{D}^2} -
\frac{\left(4 k^2 m_D^2 - \left(\pi ^2-4\right) m_D^4\right)}{4 k^2 \left(k^2+m_D^2\right)^3} \(\bm v \cdot \bm k\)^2
 \right ].
\ea
As in the case of scalar fields, the above integral contains a UV divergence independent of the Debye mass, which we remove by considering the difference
\ba
\Delta \mathcal E_{\rm EM} \equiv  \mathcal E_{\rm EM}- \mathcal E_{\rm EM} \big |_{m_D = 0}.
\ea
For this quantity, we straightforwardly find
\ba
\Delta \mathcal E_{\rm EM} = \delta M_{\rm rest} + \frac{1}{2} \delta M_{\rm kin} v^2,
\ea
where
\begin{subequations}
\label{classicalem}
\begin{eqnarray}
\delta M_{\rm rest} &=& - \frac{e_{\rm eff}^2 m_D}{ 8 \pi},
\\
\delta M_{\rm kin} &=& - \frac{e_{\rm eff}^2 m_D}{ 24 \pi} \left ( 1 - \frac{\pi^2}{16} \right ).
\end{eqnarray}
\end{subequations}

As previously advertised, the contributions to the rest and kinetic masses of the heavy quark coming from the electromagnetic field are not equal. From the above results, we see that the shift in the kinetic mass is smaller than the shift in the rest mass, and note that the discrepancy can be traced back to the imaginary part of the permittivity tensor, which is itself related to the presence of dissipation in time dependent electromagnetic fields. Evidently, the presence of dissipation decreases the energy stored in time-dependent electromagnetic fields.

\subsubsection{Assembling the results}

With Eqs.~(\ref{classicalscalar}) and (\ref{classicalem}) at hand, we are now finally ready to assemble our results for the thermal corrections to the dispersion relations of heavy quarks and scalars in SYM and heavy quarks in QCD. In the SYM theory, there are six light scalar fields for each gauge boson. From the Lagrangian of Eq.~(\ref{lagrangian}), we however see that the $\mathcal N = 2$ fields couple to only one of these, bringing the shifts in the rest and kinetic masses to
\begin{subequations}
\label{symshifts}
\begin{eqnarray}
\delta M_{\rm rest} &=& -\frac{g^2 C_F m_D}{8 \pi} + \frac{g^2 C_F m_s}{8 \pi} ,
\\
\delta M_{\rm kin} &=&- \frac{g^2 C_F m_D}{24 \pi} \left ( 1 - \frac{\pi^2}{16} \right )  + \frac{g^2 C_F m_s}{8 \pi}.
\end{eqnarray}
\end{subequations}
Here, we have expressed $e_\rmi{eff}$ in terms of the group theory factor $C_F$ and the full theory coupling $g$, while the thermal masses appearing in the results are given by
\ba
m_D^2&=& 2N_c g^2T^2 \,=\, 2m_s^2.
\ea

For QCD, where the gauge fields are the only light bosonic fields, the shifts in the rest and kinetic masses on the other hand read
\begin{subequations}
\label{qcdshifts}
\begin{eqnarray}
\delta M_{\rm rest} &=& - \frac{g^2 C_F m_D}{ 8 \pi},
\\
\delta M_{\rm kin} &=&- \frac{g^2 C_F m_D}{ 24 \pi} \left ( 1 - \frac{\pi^2}{16} \right ),
\end{eqnarray}
\end{subequations}
where the Debye mass $m_D$ has the value
\ba
m_D^2&=&\fr{g^2T^2}{3}\(N_c+\fr{N_l}{2}\).
\ea
As in the SYM case, we note that this result represents the leading thermal correction to the dispersion relation of a heavy particle for bare masses $M\gtrsim T/g$. For smaller mass scales, the leading order result can be read off from Eqs.~(\ref{symqresx}) and (\ref{highm1}).

\subsection{Quantum field theory \label{QFTcalculation}}

To complete the discussion from the beginning of this Section and to verify the result of the classical calculation, we will next compute the leading order thermal correction to the dispersion relation of infinitely heavy fundamental particles using the techniques of perturbative quantum field theory. We will do this only for an ${\mathcal N}=2$ test quark or scalar, for which we are lead to consider the graphs of Fig.~\ref{graphs1}.a--b with one gluon or light scalar line dressing the massive quark or scalar. The QCD case can later be trivially obtained from the ${\mathcal N}=2$ quark result by simply setting the scalar thermal mass to zero.

If we take the bare mass of the heavy quark to be infinitely large, we obviously must first take this limit and only then perform an expansion in powers of the coupling constant. A simple power counting exercise shows that for quarks and scalars with masses $M\gtrsim T/g$, the dominant contribution to the graphs originates from the region, where the momentum flowing along the light boson line is soft, $q_0\sim q \sim gT$. As explained above, this implies that we must use a gluon propagator dressed with the hard thermal loop (HTL) self energy and a light scalar propagator containing the scalar thermal mass $m_s$, leading us to the graphs of Fig.~\ref{dressedpropdiagram}.a--b.

\subsubsection*{Heavy fermion dispersion relation}

Let us first consider the heavy ${\mathcal N}=2$ fermion dispersion relation in the SYM theory, and thus the two graphs of Fig.~\ref{dressedpropdiagram}.b. Denoting the value of the first diagram by $\Sigma_\rmi{res}^\rmi{g}(P)$ and the second by $\Sigma_\rmi{res}^\rmi{s}(P)$, we obtain for the resummed heavy quark self energy
\ba
\Sigma_\rmi{res}(P)&=&\Sigma_\rmi{res}^\rmi{g}(P)+\Sigma_\rmi{res}^\rmi{s}(P),
\ea
where
\ba
\Sigma_\rmi{res}^\rmi{g}(P) &\equiv& g^2 C_F \sumint_Q \gamma_{\mu}\frac{-i (\slashed P - \slashed Q) + M}{(P-Q)^2 + M^2} \gamma_{\nu}D^{\mu\nu}(Q),\\
\Sigma_\rmi{res}^\rmi{s}(P)&\equiv&2i g^2C_F \slashed P \sumint_Q \fr{1}{((P-Q)^2 + M^2)(Q^2+m_s^2)},
\ea
and $D^{\mu\nu}$ denotes the resummed gluon propagator.

Performing now some straightforward algebra and using the fact that to leading order, the quark satisfies the Dirac equation, we can identify the leading large-$M$ term of the gluonic part of the self energy as
\ba
\Sigma_\rmi{res}^\rmi{g}(P)&=&-2ig^2C_F\sumint_Q \fr{P_{\mu}\gamma_{\nu}}{(P-Q)^2 + M^2}D^{\mu\nu}(Q). \label{sirqcd}
\ea
In Appendix A, we show that to leading order in a large-$M$ expansion, this function can be evaluated to yield
\ba
\Sigma_\rmi{res}(P)&=&2g^2C_F E(p)I_L\gamma_0, \label{sirqcd2}
\ea
where the real part of the integral $I_L$ is given by
\ba
{\rm Re}\, I_L&=&-\fr{m_D}{16\pi M}\bigg\{1-\fr{2p^2}{3M^2}\(1-\fr{\pi^2}{64}\)\bigg\}.
\ea

For the scalar contribution to the self energy, we on the other hand obtain
\ba
\Sigma_\rmi{res}^\rmi{s}(P)&=&-2i g^2C_F \slashed P \sumint_Q \fr{m_s^2}{((P-Q)^2 + M^2)Q^2(Q^2+m_s^2)}\nn
&=&-\fr{ig^2C_F}{8\pi}\fr{m_s}{M}\slashed P+{\mathcal O}\(\fr{1}{M^2}\).
\ea
Here, we have first regulated the UV divergence of the integral by subtracting from it the corresponding unresummed expression, and then used the fact that its leading large-$M$ piece originates from its simple $T=0$ limit.

Using the above results, we may now solve for the dispersion relation of the heavy quark from the poles of the dressed fermion propagator $1/(i\slashed P +M +\Sigma_\rmi{res}(P))$, which leads us to the equation
\ba
\Erg(p)^2&=&M^2+p^2+g^2C_F\Bigg\{\fr{m_s M}{4\pi} - \fr{\Erg(p)^2}{4\pi}\fr{m_D}{M}\bigg[1-\fr{2p^2}{3M^2}\(1-\fr{\pi^2}{64}\)\bigg]\Bigg\}.\label{N2fermdr}
\ea
Solving $\Erg(p)$ from here and expanding the result in the limit of small velocity, we finally obtain as the leading order rest and kinetic masses in the large $M$ limit\footnote{Note that here the energy is expressed in terms of the canonical momentum of the quark, and thus the rest and kinetic masses are defined through the effective Hamiltonian, $H = M_\mathrm{rest} + p^2/(2M_\mathrm{kin})$.}
\ba
M_{\rm rest}&=&M\bigg\{1-g^2C_F\fr{m_D-m_s}{8\pi M}\bigg\},\label{symqres1}\\
M_{\rm kin}&=&M\bigg\{1-g^2C_F\fr{m_D}{24\pi M}\bigg(1-\fr{\pi^2}{16}\bigg)+g^2C_F\fr{m_s}{8\pi M}\bigg\}.\label{symqres2}
\ea
These results are seen to be fully consistent with those of the classical computation. It is also worth noting that the imaginary part of $I_L$ does not contribute to the result at this order yet.

\subsubsection*{Heavy scalar dispersion relation}

For the case of a heavy ${\mathcal N}=2$ scalar, the diagrams we need to consider are depicted in Fig.~\ref{dressedpropdiagram}.a. In the large mass limit, these graphs lead to the self energy reading
\ba
\Pi_\rmi{res}(P)&=&-4g^2C_F\bigg\{M^2 I_s-\Erg(P)^2 I_L\bigg\},
\ea
and ultimately to a dispersion equation of the form
\ba
\Erg(p)^2&=&M^2+p^2+g^2C_F\Bigg\{\fr{m_s M}{4\pi}-\fr{\Erg(p)^2}{4\pi}\fr{m_D}{M}\bigg[1-\fr{2v^2}{3}\(1-\fr{\pi^2}{64}\)\bigg]\Bigg\}.
\ea
Comparing this to Eq.~(\ref{N2fermdr}), we observe perfect agreement with the case of a heavy ${\mathcal N}=2$ fermion. This was of course to be expected due to the unbroken supersymmetry of the theory in the limit under consideration.

\section{Discussion and conclusions}

In this paper, we have considered thermal corrections to the dispersion relations of fundamental particles both in an ${\mathcal N}=4$ Super Yang-Mills plasma coupled to an ${\mathcal N}=2$ hypermultiplet and in the quark gluon plasma (QGP). We have performed the calculations to the leading order in a weak coupling expansion, our main motivation having been the study of the nature of the excitations at various values of the bare (zero temperature) masses and three-momenta and a comparison of the results between the two theories. The fact that we have left out higher order corrections from our studies implies that our results are applicable only in the limit of very weak couplings (high temperatures in the case of QCD), and that we are not able to study the decay widths of the excitations.

Our results for the dispersion relation in different kinematical regions of the two theories can be read off from Eq.~(\ref{scalar}) for the ${\mathcal N}=2$ scalars, Eqs.~(\ref{symqres}) and (\ref{symqresx}) for the $\mathcal{N}=2$ quarks, and Eqs.~(\ref{qcdres1}) and (\ref{qcdres2}) for quarks in QCD. They are summarized in Table 1 displayed at the end of the paper. Inspecting the results and comparing them to each other, some interesting patterns can now be observed.

If the bare masses of the particles are small, the dispersion relation of the $\mathcal{N}=2$ quarks behaves in a way closely analogous to the dispersion relation of quarks in the QGP. For vanishing bare masses and soft three-momenta, the quantity is in both theories obtained from an equation of the form
\ba
2m^2p +\(\Erg\pm p\)\bigg\{m^2\log \left ( \frac{\Erg-p}{\Erg+p} \right ) \pm 2p^2\bigg\} &=&0,
\ea
where only the expression for the soft mass parameter $m\sim gT$ differs between the theories. This equation exhibits two solutions, a particle and a collective plasmino excitation, which are analyzed in some detail in Section 4. In the limit of larger three-momenta, $p\sim T$, the particle solution reduces to a simple form
\ba
\Erg^2 = p^2 + m_\infty^2,
\ea
where the asymptotic mass $m_\infty$ is related to $m$ above such that $m_\infty^2 = 2m^2$.

The above results for fermions are in stark contrast with the dispersion relation of the $\mathcal{N}=2$ scalars. For the latter, the dispersion relation is for all values of the bare mass $M$ and three-momentum $p$ given by
\ba
\Erg = \sqrt{p^2 + M^2 + \delta M_s^2},
\ea
where $\delta M_s^2$ is independent of the momentum $p$ but depends on the bare mass $M$ as shown in Fig.~2. We thus see that for soft bare masses and three-momenta, the behavior of the dispersion relations in the SYM theory does not exhibit supersymmetry: The fermions and scalars within the same SUSY multiplet behave differently. This can be attributed to the presence of a non-zero temperature in the system, which is known to explicitly violate SUSY.

Approaching bare masses of order $T$, the dispersion relation of the quarks can in both theories be written in the form
\ba
\Erg = \sqrt{p^2 +M^2+\delta M_q^2}.
\ea
Here, the mass correction $\delta M_q^2$ is equal to the scalar mass correction $\delta M_s^2$ for the $\mathcal{N}=2$ quarks, while for quarks in QCD, it becomes a slowly varying function of $p$ that we displayed in Fig.~4. We thus observe that the dispersion relations of the ${\mathcal N}=2$ quarks and scalars become equal in this limit, and that supersymmetry is restored at least in this sense. The same is true, if the bare mass is small but the three-momentum hard, since $\delta M_s$ approaches $m_\infty$ in this limit.

The above observations are in agreement with the conclusions drawn in Ref.~\cite{CaronHuot:2008uw}, and can be easily understood in the limit, where either $M$ or $p$ becomes the dominant scale in the problem. Although the heat bath explicitly breaks SUSY, the breaking of the symmetry is not manifest in the propagation of the particles unless they are sufficiently light and soft, as heavy $\mathcal{N}=2$ propagators act in the diagrams essentially as zero-temperature ones. For fundamental particle masses $M\gtrsim T$, the leading thermal contributions to the self energies in fact come from the screening of the $\mathcal{N}=4$ fields, to which the fields in the $\mathcal{N}=2$ hypermultiplets couple to identically, with couplings independent of $T$. What we find to be somewhat surprising is that in our calculations this effect is seen to take place as soon as $M$ or $p$ is parametrically larger than $gT$, and it is only strictly in the limit of soft momenta and bare masses that we see any difference between the ${\mathcal N}=2$ fermions and scalars.

Moving on to yet larger masses, a particularly interesting special case of our results is encountered, when the bare mass of the fundamental fields becomes parametrically larger than the temperature. In this limit, we may compare our results to the findings of Ref.~\cite{hky}, where the authors
considered the non-relativistic dispersion relation of heavy ($M\gg \sqrt{\lambda}T = \sqrt{N_c}\,gT$) ${\mathcal N}=2$ quarks and scalars in the large $N_c$ limit of a strongly coupled SYM plasma. According to their results, the excitations follow a dispersion relation of the form
\ba
\Erg &=& M_\rmi{\rm rest}(T)+\fr{p^2}{2M_\rmi{\rm kin}(T)},
\ea
where both $M_\rmi{\rm rest}$ and $M_\rmi{\rm kin}$ can in the limit of a large rest mass be given in the form
\ba
M_\rmi{T}&=&M\Bigg\{1-\fr{\widetilde{m}}{M}+{\mathcal O}\(\fr{\widetilde{m}^2}{M^2}\)\Bigg\},\\
\widetilde{m}&\equiv&\fr{1}{2}\sqrt{\lambda}T,
\ea
and the difference between $M_\rmi{\rm rest}$ and $M_\rmi{\rm kin}$ is suppressed by at least one inverse power of $M$ and is displayed numerically in Fig.~1 of Ref.~\cite{hky}.

This behavior is to be contrasted with our results from Section~\ref{nonrellimit}, which state that for heavy non-relativistic ${\mathcal N}=2$ quarks and scalars immersed in a weakly coupled SYM plasma, the shifts in the rest and kinetic masses are given by Eq.~(\ref{symshifts}). Going to the large $N_c$ limit and defining
\ba
\bar{m}&\equiv&\fr{\lambda^{3/2}T}{8\pi},
\ea
we obtain from Eq.~(\ref{symshifts})
\ba
M_{\rm rest}&=&M\left \{1-(\sqrt{2}-1)\fr{\bar{m}}{M}\right\} \approx M\left \{1-0.4 \fr{\bar{m}}{M}\right\},\\
M_{\rm kin}&=&M\left \{1+\(1-\fr{\sqrt{2}(1-\pi^2/16)}{3}\)\fr{\bar{m}}{M}\right \} \approx M\left \{1+0.8 \fr{\bar{m}}{M}\right\}.
\ea
We observe that the sign of the shift in the rest mass is the same both in the weakly and strongly coupled limits: In both cases thermal effects lower the value of the mass. However, the signs of the shifts in the kinetic mass do not agree. In the weakly coupled limit, the contribution to the quantity from gauge fields is negative, but smaller in magnitude than the effect of the scalar fields, which gives the shift a positive overall sign. Moreover, in the weakly coupled limit the difference $M_\mathrm{rest}-M_\mathrm{kin}$ is constant and non-zero as $M\rightarrow \infty$, in contrast to the result in the strongly coupled limit.

Some interesting observations can now be made. Although the shifts in the effective rest and kinetic masses are not equal in the weakly coupled limit, the scalar contributions to them are, implying that the difference in the masses arises from the dynamics of the gauge fields alone. At leading order in a weak coupling expansion, a physical difference between the scalar and gauge field dynamics is that the former is not dissipative. Indeed, it is easily seen that a non-zero difference between the kinetic and rest masses requires a momentum dependent permittivity tensor.  Moreover, by the Kramers-Kronig relations, any momentum dependence in the real part of the permittivity tensor implies that its imaginary part must be non-zero.  At least at weak coupling, we therefore see that there is a connection between having a non-vanishing difference between the rest and kinetic masses and the dissipation of gauge fields. While this argument need not hold at larger coupling, it is interesting to contrast our observation with the fact that the mass difference is suppressed by $1/M$ in strongly coupled SYM theory. We leave the exploration of this subject for future work.

\section*{Acknowledgments}

We are indebted to Simon Caron-Huot and Larry Yaffe for their valuable advice regarding determining the dispersion relation of non-relativistic particles, and in addition wish to thank Francois Gelis, Mikko Laine and Dam Son for useful discussions. P.C.~was supported by the U.S.~Department of Energy under Grant No.~DE-FG02-96ER40956, A.G.~by the Austrian Science Foundation FWF, project no. P19526-N16, and A.V.~by the Austrian Science Foundation FWF, project No.~M1006, as well as the Sofja Kovalevskaja Award of the Humboldt foundation.

\appendix

\section{Evaluation of $\Sigma_\rmi{res}^\rmi{g}$}

In this Appendix, we evaluate the large-$M$ limit of the sum-integral defined by Eq.~(\ref{sirqcd}). It is needed in determining the leading order thermal correction to the heavy quark dispersion relation due its interaction with a dressed gluon, and is relevant in both SYM and QCD. We will perform the calculation in an expansion in powers of the velocity $v$ up to and including order $v^2$, as in Section 6 we are ultimately only interested in the thermal shifts to the rest and kinetic masses of the quark.

Taking advantage of the gauge invariance of the quantity under consideration, we choose to work in the Coulomb gauge, in which the dressed Euclidean space gluon propagator reads
\ba
D^{\mu\nu}(Q) &=& \fr{\delta_{\mu 0}\delta_{\nu 0}}{q^2+\Pi_{00}}+\fr{\delta_{ij}-\hat{q}_i\hat{q}_j}{q^2+q_0^2+\Pi_\rmi{T}},
\ea
with the components of the HTL polarization tensor having the forms
\ba
\Pi_{00}(Q)&=&m_D^2\bigg\{1+\fr{iq_0}{2q}\ln\,\fr{iq_0-q}{iq_0+q}\bigg\},\\
\Pi_\rmi{T}(Q)&=&-m_D^2\bigg\{\fr{q_0^2}{2q^2}+\fr{iq_0(q_0^2+q^2)}{4q^3}\ln\,\fr{iq_0-q}{iq_0+q}\bigg\}
\ea
and $m_D$ denoting the Debye mass of the theory. We also choose the branch cut of the logarithm to be slightly tilted off from the negative real axis, so that these expressions have a branch cut going from $q_0=-iq$ to $q_0=iq$ via the left side of the complex plane (see Fig.~\ref{figcp1}). This is required in order to keep the polarization tensor an even function of $q_0$ when evaluated at real Matsubara frequencies, while maintaining the contribution of the small-$|q_0|$ limit well-defined.

Having the vector $\bm p$ point in the $z$ direction of our coordinate system, it is easy to see that the above produces
\ba
\!\!\!\!\!\!\Sigma^g_\rmi{res}(P)&=&-2ig^2C_F\sumint_Q \fr{1}{(P-Q)^2 + M^2}\bigg(\fr{1}{q^2+\Pi_{00}}p_0\gamma_0+\fr{1-\hat{q}_z^2}{q_0^2+q^2+\Pi_\rmi{T}}p_i\gamma_i\bigg).
\ea
We regulate the UV divergences appearing here by subtracting from the integrand the corresponding unresummed expression, which in addition helps us to isolate the leading large-$M$ limit of the graph we are interested in.\footnote{It is easy to confirm that for the unresummed integral, the finite temperature contribution is suppressed by two inverse powers of the heavy mass, while the $T=0$ piece is part of the zero temperature quantum corrections we are not interested in. We are thus justified to make the subtraction.} This gives us
\ba
\Sigma^\rmi{reg}_\rmi{res}(P)&=&-2ig^2C_F\sumint_Q \fr{1}{(P-Q)^2 + M^2}\bigg(\bigg[\fr{1}{q^2+\Pi_{00}}-\fr{1}{q^2}\bigg]p_0\gamma_0\nn
&+&\bigg[\fr{1-\hat{q}_z^2}{q_0^2+q^2+\Pi_\rmi{T}}
-\fr{1-\hat{q}_z^2}{q_0^2+q^2}\bigg]p_i\gamma_i\bigg)\nn
&\equiv&-2ig^2C_F\bigg\{I_L p_0\gamma_0+I_T p_i\gamma_i\bigg\},
\ea
which defines the longitudinal and transverse integrals, $I_L$ and $I_T$, that we now set out to evaluate. As the imaginary parts of the integrals only affect the dispersion relation at higher orders in $g$ than what we are pursuing in this work, below we will only consider their real parts.

\subsection{The longitudinal integral}

\begin{figure}[t]

\centerline{\epsfxsize=10.3cm\epsfysize=5.3cm \epsfbox{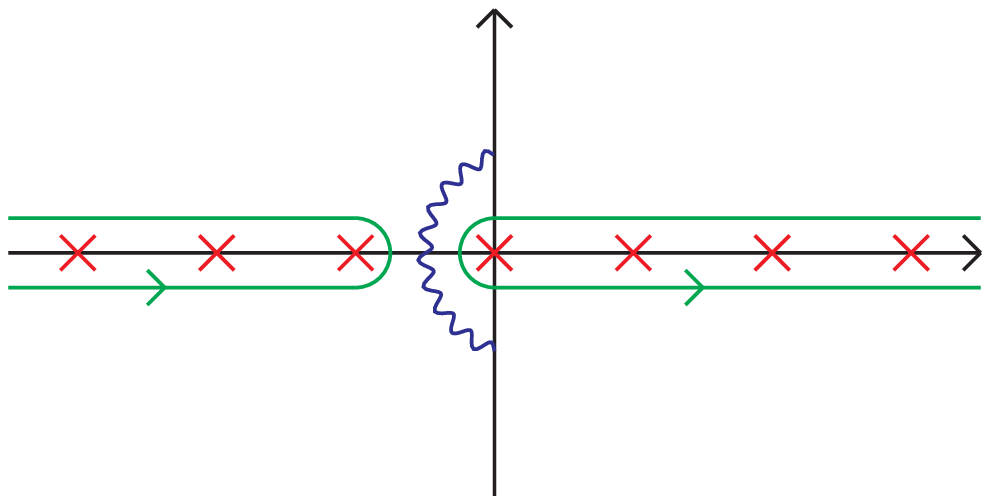}}

\caption[a]{Illustration of the contour $C$ on the complex $q_0$ plane. The red crosses denote the locations of the bosonic Matsubara modes at $q_0=2n\pi T$, the blue wiggly line the branch cut of the integrand going from $q_0=-iq$ to $q_0=iq$, and the green curve our integration contour. The contour $C$ encircles all the poles of the cotangent function, yet avoiding crossing the branch cut.} \label{figcp1}
\end{figure}

Let us first look at the longitudinal case. The Matsubara sum appearing in the definition of the integral can be transformed into a contour integral  using standard tricks, but recalling that there is now a branch cut in the definition of the polarization tensor that the integration path should avoid. Choosing thus a somewhat non-standard contour $C$ according to Fig.~\ref{figcp1} and defining $g(q_0)\equiv\fr{1}{(P-Q)^2 + M^2}(\fr{1}{q^2+\Pi_{00}}-\fr{1}{q^2})$, we obtain
\ba
I_L&=&\int\!\! \fr{d^3q}{(2\pi)^3}\oint_C \fr{dq_0}{4\pi} g(q_0)\cot \fr{\beta q_0}{2}\nn
&\simeq&\fr{1}{4\pi}\int\!\! \fr{d^3q}{(2\pi)^3}\bigg\{\int_{-\infty}^{-(c+\e)}dq_0\(g(q_0+i\e)+g(q_0-i\e)\)\nn
&+&\int_{-(c-\e)}^{\infty}dq_0\(g(q_0+i\e)+g(q_0-i\e)\)\bigg\}\nn
&\simeq&\fr{1}{4\pi}\int\!\! \fr{d^3q}{(2\pi)^3}\int_{-\infty}^{\infty}dq_0\(g(q_0+i\e)+g(q_0-i\e)\)\nn
&=&\int\!\! \fr{d^3q}{(2\pi)^3}\int_{-\infty}^{\infty}\fr{dq_0}{2\pi}g(q_0). \label{intcont1}
\ea
Here, we have first neglected the part of the cotangent function that vanishes in the $T\rightarrow 0$ limit\footnote{On dimensional grounds, this is again suppressed by at least two powers of $M$.}, then denoted the point at which the branch cut intersects the real axis by $-c$, and finally used the fact that the function $g$ is analytic on the Euclidean real axis to proceed to the $\e\rightarrow 0$ limit. In the last form, we are integrating the function $g$ through the branch cut, which is obviously the correct prescription in this case. It will be seen later that the location of the branch cut --- as long as it doesn't lie on the imaginary axis --- does not affect the final result of the integral.

Writing now the integral in its full form, we have
\ba
I_L&=& \int\!\!\fr{d^4q}{(2\pi)^4}\fr{1}{(p_0-q_0)^2+(\bm p-\bm q)^2+M^2}\bigg\{\fr{1}{q^2+\Pi_{00}(q_0,q)}-\fr{1}{q^2}\bigg\}\nn
&=& -\int\!\!\fr{d^4q}{(2\pi)^4}\fr{1}{q_0^2+q^2-2iq_0\sqrt{M^2+p^2}-2\bm q\cdot\bm p}\fr{\Pi_{00}(q_0,q)}{q^2(q^2+\Pi_{00}(q_0,q))}, \label{IL1}
\ea
where in the latter form we have already taken the on shell limit of the heavy particle, $p_0=iE(p)$. This implies that in order to correctly perform the analytic continuation in the integral, we must slightly deform the $q_0$ integration contour, so that it goes between the two poles of the fermionic propagator, as it did for real $p_0$.

To further analyze the integral of Eq.~(\ref{IL1}), we note that if either $q_0$ or $q$ was parametrically larger than $m_D$, (\textit{e.g.~}of order $M$), then it would be easy to see that the integral would scale as $(m_D/M)^{1+\alpha}$, with $\alpha>0$, \textit{i.e.~}be subleading with respect to the contributions we expect to find. We will thus rescale the integration variables according to
\ba
x&\equiv&\fr{m_D}{M},\\
\bm q&\equiv& m_D \bm k,\\
q_0&\equiv&m_Dkz,\\
\Pi_{00}(q_0/q)&\equiv &m_D^2f(z)\\
\bm p&\equiv& M\bm v,
\ea
in terms of which we have
\ba
I_L= -x\int_{C'}\fr{dz}{2\pi}\int\!\!\fr{d^3k}{(2\pi)^3}\fr{1}{k}\fr{1}{x(1+z^2)k^2-2izk\sqrt{1+v^2}-2\bm v\cdot\bm k}\fr{f(z)}{k^2+f(z)},\label{i1}
\ea
with $C'$ reminding us of the deformation of the $q_0$ integration path. Note that in our notation throughout this Appendix, $\bm v$ just a shorthand for $\bm p/M$, and not the correct relativistic velocity of the heavy particle. This is not a problem, as we are merely interested in the two first terms of the small-$v$ expansion of the dispersion relation.

\begin{figure}[t]

\centerline{\epsfxsize=7.3cm\epsfysize=5.3cm \epsfbox{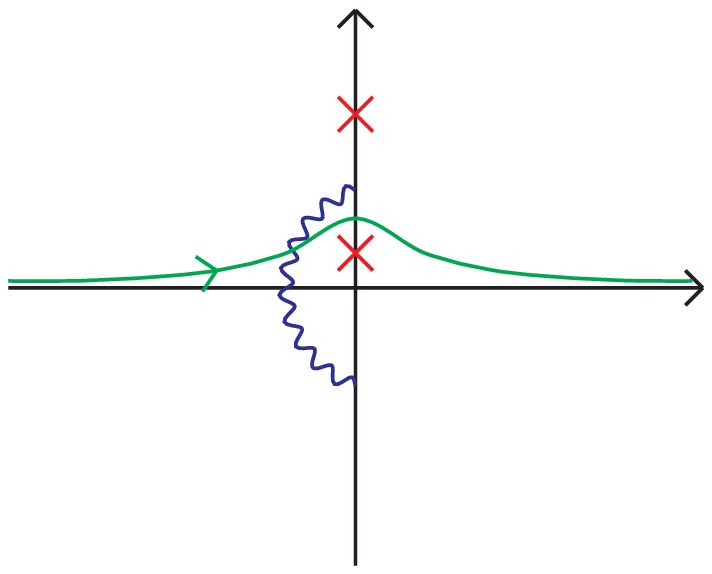}}

\caption[a]{Illustration of the contour $C'$. The red crosses denote here the two imaginary poles of the fermion propagator, but otherwise the notation is as in Fig.~\ref{figcp1}. \label{figcp}}
\end{figure}

Next, we point out that out of the two poles of the fermion propagator
\ba
z&=&z_{\pm}\;\equiv\;\fr{i}{xk}\(\sqrt{1+v^2}\pm\sqrt{1+(\bm v-x\bm k)^2}\),
\ea
both are imaginary, one being large ($\gg i$) and always on the upper half plane, while the other can reside on either side of the real axis, depending on $\bm k$, but is infinitesimally small in magnitude ($\sim x$ or $v$). They are marked with red crosses in Fig.~\ref{figcp}. As before, the dressed gluon propagator has a branch cut (marked in blue) running from $z=-i$ to $z=i$ that is slightly deformed from the imaginary axis. As noted above, in order for the analytic continuation to the mass shell limit of the heavy particle to have been performed correctly, the $z$ integration contour (green) has to be chosen to go in between the two poles of the heavy field propagator and cross the imaginary axis between $z=z_{-}$ and $z=i$. Outside the vicinity of the origin of the complex plane, we may choose $C'$ to lie on the real axis.

For concreteness, let us now define $C'$ in the following way: It follows the real axis all the way from $z=-\infty$ to the origin, then --- if and only if $z_{-}$ resides on the upper half of the complex plane --- makes a detour shooting up along the imaginary axis to encircle the pole at $z= z_{-}$ in the clockwise direction and returning to the origin, and finally proceeds along the real axis to $z=+\infty$. As the branch cut of the gluon propagator has been shifted away from the imaginary axis, the contributions from going up and down the imaginary axis cancel, and we are left with two separate parts: One from the integral along the real axis, and another from encircling the pole at $z=z_{-}$ in the clockwise direction. We denote these two functions by $I_L^{\rm real}$ and $I_L^{\rm pole}$, respectively, and now proceed to expand them in a power series in $v$ while working only to leading order in small $x$. One possible non-triviality in this procedure is related to the order of taking $v$ and $x$ to zero, which we have dealt with by separately considering the cases $v\sim x^0$ and $v\sim x$, and checking that they lead to consistent results. Below we will, however, only present the calculations in the former limit, which corresponds to first taking the small-$x$ and only then the small-$v$ limit.

\subsubsection*{The pole contribution}

Let us first inspect the pole contribution, $I_L^{\rm pole}$, which is only present for momenta $\bm k$ that satisfy $\bm v\cdot\hat{\bm k} > x k/2$. Denoting the angle between $\bm v$ and $\bm k$ by $\theta$, we easily obtain
\ba
I_L^{\rm pole} &=& -\fr{x}{8\pi^2}\int_0^\infty {\rm d}k \int_0^\pi {\rm d}\theta \sin\theta \fr{\Theta(\cos\theta -\fr{xk}{2v})}{\sqrt{1+(\bm v-x\bm k)^2}}\fr{f(z_{-})}{k^2+f(z_{-})}\nn
&\equiv&-\fr{x}{8\pi^2}F(x,v),
\ea
where for $v$ non-zero and of order $x^0$, we may set $x$ to zero in $F$. This leads to
\ba
F(0,v) &=& \int_0^\infty {\rm d}k \int_0^{\pi/2} {\rm d}\theta \sin\theta
\fr{1}{\sqrt{1+v^2}}
\fr{f(i v \cos\theta)}{k^2+f(i v \cos\theta)},
\ea
in which we expand $f$ in powers of $v$, using the fact that consistently with our choice of the branch cut, we have around the origin
\ba
f(z)&=&1+\fr{iz}{2}\bigg(\log\,\fr{1-iz}{1+iz}+i\pi\bigg).
\ea
This enables us to perform the above integrations with ease, giving for the real part of $I_L^{\rm pole}$
\ba
{\rm Re}\, I_L^{\rm pole} &=& -\fr{x}{16\pi}\bigg\{1-\fr{2v^2}{3}\(1-\fr{\pi^2}{64}\) + {\mathcal O}(v^4)\bigg\}.
\ea

\subsubsection*{The real axis contribution}

Next, we move on to consider the integral along the real axis. Here, we take advantage of the reality of $z$ and write $I_L$ in the form
\ba
I_L^{\rm real}&=& -x\int_{-\infty}^\infty\fr{dz}{2\pi}\int\!\!\fr{d^3k}{(2\pi)^3}\fr{1}{k}\fr{f(z)}{k^2+f(z)}
\bigg\{\fr{x(1+z^2)k^2-2\bm v\cdot\bm k}{(x(1+z^2)k^2-2\bm v\cdot\bm k)^2+4z^2k^2(1+v^2)}\nn
&+&i\,\fr{2zk\sqrt{1+v^2}}{(x(1+z^2)k^2-2\bm v\cdot\bm k)^2+4z^2k^2(1+v^2)}\bigg\},
\ea
where the latter, imaginary part obviously vanishes due to its antisymmetricity in $z$. For the real part, we on the other hand see that for $z$ of order $x^0$, the $\bm k$ integral vanishes by symmetry to leading order in $x$, so we must let $z=x\widetilde{z}$. Setting then $x\rightarrow 0$ wherever possible, we obtain in terms of this new variable
\ba
\!\!\!\! I_L^{\rm real}&=&-x\int_{-\infty}^\infty\fr{d\widetilde{z}}{2\pi}\int\!\!\fr{d^3k}{(2\pi)^3}\fr{1}{k}\fr{f(0)}{k^2+f(0)}
\fr{k^2-2\bm v\cdot\bm k/x}{(k^2-2\bm v\cdot\bm k/x)^2+4\widetilde{z}^2k^2(1+v^2)}.
\ea
This expression is obviously at least ${\mathcal O}(x^2)$, if $v$ is of order $x^0$, and we may hence neglect it here.

\subsubsection*{Final result}

Assembling the two pieces of our calculation together, we finally obtain the desired result
\ba
{\rm Re}\, I_L &=&{\rm Re} \(I_L^{\rm pole}+I_L^{\rm real}\)\; = \; -\fr{x}{16\pi}\bigg\{1-\fr{2v^2}{3}\(1-\fr{\pi^2}{64}\)+ {\mathcal O}(v^4)\bigg\}.
\ea
This result may be verified by considering velocities $v$ of order $x$, in which case the individual results for $I_L^{\rm pole}$ and $I_L^{\rm real}$ however differ from the above expressions.

\subsection{The transverse integral}

With the transverse integral, $I_T$, our task is somewhat simpler, as to obtain the rest and kinetic masses, it suffices to consider the integral in the zero external three-momentum limit, $\bm v =0$. The calculation proceeds in almost exact parallel with the above one, with the only subtlety being related to the unphysical Landau pole $q_0=q_L(q)$ that appears when evaluating the transverse propagator on the Euclidean real axis between the branch cut and $q_0=0$.\footnote{Solving for the poles of the transverse gluon propagator between the branch cut and the imaginary axis, one finds the so-called Landau pole at $q_0=q_L(q)\approx-\fr{4q^3}{\pi m_D^2}$, where the last approximation corresponds to the $q\rightarrow 0$ limit.} Due to this pole, one cannot combine the $q_0$-integration as in going from Fig.~\ref{figcp1} to Eq.~(\ref{intcont1}), but must evaluate it in two pieces, from $q_0=-\infty$ to the branch cut and from $q_0=q_L(q)+\e$ to infinity.

Denoting the integration path described above by $C^{''}$, it is straightforward to see that to leading order in $x$ we obtain
\ba
I_T&\equiv& \sumint_Q \fr{1-\hat{q}_z^2}{(P-Q)^2 + M^2}\bigg\{\fr{1}{q_0^2+q^2+\Pi_\rmi{T}}-\fr{1}{q_0^2+q^2}\bigg\}\nn
&=&-\fr{ix}{2}\int_{C^{''}}\fr{dz}{2\pi}\fr{1}{z(1+z^2)}\int\!\!\fr{d^3k}{(2\pi)^3}\fr{1-\hat{k}_z^2}{k^2} \fr{\widetilde{f}(z)}{(1+z^2)k^2+\widetilde{f}(z)},\label{it2}
\ea
where in the vicinity of the origin, $z=0$, the function $\widetilde{f}(z)$ is given by
\ba
\widetilde{f}(z)&=&-\fr{z^2}{2}-\fr{iz(1+z^2)}{4}\bigg(\log\,\fr{1-iz}{1+iz}+i\pi\bigg).
\ea
Observing that the integrand in Eq.~(\ref{it2}) is imaginary everywhere on the Euclidean real axis, we conclude that all real contributions originate from avoiding the possible pole at $z=0$. This time it, however, is immediately obvious from the form of $\widetilde{f}(z)$ that the corresponding residue vanishes (\textit{i.e.~}there in fact is no pole at $z=0$), from
which it follows that
\ba
I_T&=&\mathcal O(v^2).
\ea

\subsection{The self energy}

Assembling everything together, we finally see that we obtain the result of Eq.~(\ref{sirqcd2}) for the self energy function $\Sigma_\rmi{res}^\rmi{g}(P)$, evaluated to leading order in $x=m_D/M$ and to order $v^2$ in powers of the velocity.

\section{Integral representations of sum-integrals \label{sumints}}

Here, we assemble the integral representations of various sum-integrals needed in our calculations. Using standard integration tricks, one easily obtains
\ba
\label{irel}
&&\sumint_{\pm} \frac{1}{(P+K)^2 + M^2} \frac{1}{K^2}  \nn
&=& \pm\frac{1}{8 \pi^2p}\bigg\{\int_{0}^{\infty}  dk\;\real\left[\log \left (\frac{P^2 + M^2 + 2 i k p_0 + 2 p k}{ P^2 + M^2 + 2 i k p_0 - 2 p k} \right )\right] \frac{1}{e^{\beta k} \mp 1} \nn \label{B3}
&-& \int _{0}^{\infty} dk \frac{k}{E(k)} \real\left[\log \left ( \frac{-P^2 + M^2 +2 i E(k) p_0 + 2 p k}{-P^2 + M^2 +2 i E(k) p_0 - 2 p k} \right )\right]  \frac{\mathrm{sgn}(P)}{e^{\beta E(k)} \mp \mathrm{sgn}(P)}\bigg\},  \\
&&\sumint_{\pm} \frac{1}{K^2+M^2} \;=\; \pm \frac{1}{2 \pi^2} \int_{0}^{\infty} dk \frac{k^2}{E(k)} \frac{1}{e^{\beta E(k)} \mp 1}\label{B2},
\ea
where $\mathrm{sgn}(P) = 1$ for bosonic and $-1$ for fermionic $P$, $E(k)=\sqrt{k^2+M^2}$, and the operation $\real$ is defined by $\real\,f(p_0) = 1/2\,[f(p_0)+f(-p_0)]$.


\begin{table}
\begin{tabular}{l|ll}
 Kinematic region & \multicolumn{2}{|l}{Dispersion relation}  \\
\hline
$M = 0,\, p\lesssim gT$& QCD: & \\
 & &
 $
  2\widetilde{m}^2p + (\mathcal{E}\pm p)\left\{\widetilde{m}^2\,\mathrm{log}\displaystyle \frac{\mathcal{E}-p}{\mathcal{E}+p}\pm 2p^2 \right\} = 0,
 $
 \\
 & &
 $ \widetilde{m}^2 = \displaystyle \frac{g^2 C_F T^2}{8} $  \\
 & SUSY: &  \\
& quarks:&
 $
  2m^2p + (\mathcal{E}\pm p)\left\{m^2\,\mathrm{log}\displaystyle \frac{\mathcal{E}-p}{\mathcal{E}+p}\pm 2p^2 \right\} = 0
 $
 \\
& &
 $ m^2 = \displaystyle \frac{g^2 C_F T^2}{2} $  \\
 & &  \\
& scalars: &
 $ \mathcal{E}^2 = p^2 + \delta M_s^2, \hspace{0.5cm} \delta M_s^2 = g^2 C_F T^2$  \\
\hline
$M = 0,\, p\gg gT$& QCD: & \\
 & &
 $ \mathcal{E}^2 = p^2 + m^2,\hspace{0.5cm} m^2 = \displaystyle \frac{g^2 C_F T^2}{4} $  \\
& SUSY: & \\
& &
 $ \mathcal{E}^2 = p^2 + \delta M^2, \hspace{0.5cm} \delta M^2 = g^2 C_F T^2$  \\
\hline
$gT \ll M \ll T/g, $ & QCD: & (see Fig.~4) \\
$ p $ unconstrained & &
 $ \mathcal{E}^2 = p^2 + M^2 + \delta M^2,\hspace{0.5cm} \delta M = \delta M(p,M)$  \\
 & SUSY: & (see Fig.~2) \\
& & $ \mathcal{E}^2 = p^2 + M^2 + \delta M^2, \hspace{0.5cm} \delta M = \delta M(M) $   \\
\hline
$ M \gtrsim T/g, $ & QCD: & \\
$p$ non-relativistic & & $\mathcal{E} = M_\mathrm{rest} + \displaystyle \frac{1}{2}M_\mathrm{kin}v^2,$  \\
 & & $M_\mathrm{rest} = M - \displaystyle \frac{g^2 C_F m_D}{8\pi}$ \\
$m_D$ and $m_s$ refer & & $M_\mathrm{kin} = M - \displaystyle{\frac{g^2 C_F m_D}{24\pi}\left(1-\frac{\pi^2}{16} \right)}$ \\
to the static & SUSY: &  \\
thermal screening & & $\mathcal{E} = M_\mathrm{rest} + \displaystyle \frac{1}{2}M_\mathrm{kin}v^2,$ \\
masses of the  & & $M_\mathrm{rest} = M - \displaystyle{\frac{g^2 C_F m_D}{8\pi} + \frac{g^2 C_F m_s}{8\pi}}$ \\
gluons and scalars & & $M_\mathrm{kin} = M - \displaystyle{\frac{g^2 C_F m_D}{24\pi}\left(1-\frac{\pi^2}{16}\right) + \frac{g^2 C_F m_s}{8\pi}}$ \\
\hline
\end{tabular}
\caption{This table summarizes our results. The SUSY results apply for both the fundamental quarks and scalars in the $\mathcal{N}=2$ hypermultiplet unless otherwise stated.}
\label{summarytable}
\end{table}

\end{document}